\documentclass[iop,twocolappendix,numberedappendix,appendixfloats]{emulateapj}
\newcommand{\Msun}{\,M_{\odot}}

\newcommand{\be}{\begin{equation}}
\newcommand{\ee}{\end{equation}}

\def\teff{T_{\rm eff}}

\def\lbol{L_{\rm bol}}
\def\HST{\emph{HST}}
\def\Inorm{I_{\rm rms,norm}}
\def\Irms{I_{\rm rms}}
\def\Vrms{V_{\rm rms}}
\def\I{I_{\rm 814}}
\def\V{V_{\rm 606}}
\def\actaa{Act. Astron.}

\shortauthors{CONROY ET AL.}

\shorttitle{Stellar Variability on Day to Decade Timescales}

\begin{document}


\title{A Complete Census of Luminous Stellar Variability on Day to
  Decade Timescales}

\author{Charlie Conroy\altaffilmark{1}, Jay Strader\altaffilmark{2},
  Pieter van Dokkum\altaffilmark{3}, Andrew E. Dolphin\altaffilmark{4},
  Daniel R. Weisz\altaffilmark{5}, Jeremiah W. Murphy\altaffilmark{6}, Aaron
  Dotter\altaffilmark{1}, Benjamin D. Johnson\altaffilmark{1}, Phillip
  Cargile\altaffilmark{1}}

\altaffiltext{1}{Department of Astronomy, Harvard University,
  Cambridge, MA, 02138, USA}
\altaffiltext{2}{Department of Physics and Astronomy, Michigan State
  University, 567 Wilson Road, East Lansing, MI 48824, USA}
\altaffiltext{3}{Department of Astronomy, Yale University, New Haven,
 CT, 06511, USA}
\altaffiltext{4}{Raytheon Company, Tucson, AZ, 85734, USA}
\altaffiltext{5}{Department of Astronomy, University of California
  Berkeley, Berkeley, CA 94720, USA}
\altaffiltext{6}{Department of Physics, Florida State University, 600
  W College Avenue, Tallahassee, FL 32306, USA}

\slugcomment{Submitted to ApJ}

\begin{abstract}

  Stellar photometric variability offers a novel probe of the interior
  structure and evolutionary state of stars.  Here we present a census
  of stellar variability on day to decade timescales across the
  color-magnitude diagram for 73,000 stars brighter than
  $M_{\rm I,814}=-5$ in the Whirlpool Galaxy (M51).  Our Cycle 24
  \HST\, program acquired $\V$ and $\I-$band images over 34 epochs
  spanning one year with pseudo-random cadences enabling sensitivity
  to periods from days to months.  We supplement these data with
  archival $V$ and $I-$band \HST\, data obtained in 1995 and 2005,
  providing sensitivity to variability on decade timescales.  At least
  50\% of stars brighter than $M_{\rm I,814}=-7$ show strong evidence
  for variability within our Cycle 24 data; amongst stars with
  $\V-\I>2$ the variability fraction rises to $\approx100$\%.  Large
  amplitude variability ($>0.3$ mag) on decade timescales is
  restricted to red supergiants and very luminous blue stars.  Both
  populations display fairly smooth variability on month-year
  timescales.  The Cepheid instability strip is clearly visible in our
  data, although the variability fraction within this region never
  exceeds $\approx10$\%.  The location of variable stars across the
  color magnitude diagram broadly agrees with theoretical sources of
  variability, including the instability strip, red supergiant
  pulsational instabilities, long-period fundamental mode pulsations,
  and radiation-dominated envelopes in massive stars.  Our data can be
  used to place stringent constraints on the precise onset of these
  various instabilities and their lifetimes and growth rates.

\end{abstract}

\keywords{stars: variables: general, Hertzsprung-Russell and
  colour-magnitude, galaxies: individual (NGC 5194)}


\section{Introduction}
\label{s:intro}

Stellar variability is a powerful tool to study a variety of phenomena
in the Universe, including stellar interiors (via asteroseismology),
the final stages of massive stars (via eruptive behavior, e.g., $\eta$
Carinae), the cosmic distance ladder (via the ``Leavitt Law"---the
Cepheid period--luminosity relation), stellar masses and radii (via
eclipsing binaries and period--luminosity relations of evolved stars),
and the explosive behavior of novae and supernovae \citep[see reviews
in][]{Eyer08, Catelan15}.

The timescales, regularity, and amplitudes of stellar variability span
essentially the entire observable parameter space. Variability occurs
on the shortest observed timescales (seconds to minutes in the case of
helioseismology) and the longest probed timescales (the light curves
for $\eta$Car and P Cyg span $\sim400$ years). Some variables show
regular, well-defined periods over long timescales, while others have
unpredictable irregular behavior, such as eruptions of luminous blue
variables (LBVs), in addition to the more spectacular novae and
supernovae. Finally, variability amplitudes range from the smallest
detectable fluctuations \citep[ppm; e.g.,][]{Borucki10}, to several
magnitude fluctuations amongst the LBVs, red supergiants (RSGs), and
Mira variables, to much larger amplitudes for the novae and
supernovae.

Stellar variability can be driven by a variety of physical processes.
The best understood on theoretical grounds are the excitation of
unstable radial or non-radial modes, as encountered in
asteroseismology, luminous pulsating red variables such as Miras and
semi-regular variables (SRVs) \citep[e.g.,][]{Wood79}, Cepheids, RV
Tau stars, and many others.  These stars tend to experience moderate
amplitude variability in their physical properties and for these
reasons the variability is rarely destructive to the star.  On the
other hand, large amplitude variability, such as occurs in RSGs and
LBVs can alter and/or hasten the subsequent evolution of the star
\citep{Heger97, Yoon10, Owocki15, Smith06, Smith14c}.  Evolution of
very massive stars ($>20\Msun$) is quite uncertain both because of
limited observational samples and because the underlying physics is
poorly understood.  For example, the physical origin of the luminosity
variation observed in LBVs is a mystery, although there are many
possibilities including violent strange mode instabilities
\citep{Yadav17}, unstable turbulent convection \citep{Smith14b},
binary star merger products \citep{Smith14b, Justham14}, and
radiation-dominated stellar envelopes near the local Eddington limit
\citep{Paxton13, Jiang15}.  Progress is limited in part by the rarity
of massive stars and the long timescales involved, both of which make
it challenging to test models against observations.

There have been many surveys of stellar variability in the Local Group
and beyond, dating back to the early part of the last century
\citep[e.g.,][]{Leavitt08, Hubble53, Tammann68}.  Time domain surveys
have particularly proliferated in the past twenty years, with the
advent of fast, wide format imaging systems. Examples include MACHO
\citep{Alcock97}, ASAS \citep{Pojmanski02}, OGLE \citep{Udalski03},
Pan-STARRS1 \citep{Chambers16}, PTF \citep{Law09}, CRTS
\citep{Drake14}, and ASAS-SN \citep{Jayasinghe18}.

\begin{figure}[!t]
\center
\includegraphics[width=0.4\textwidth]{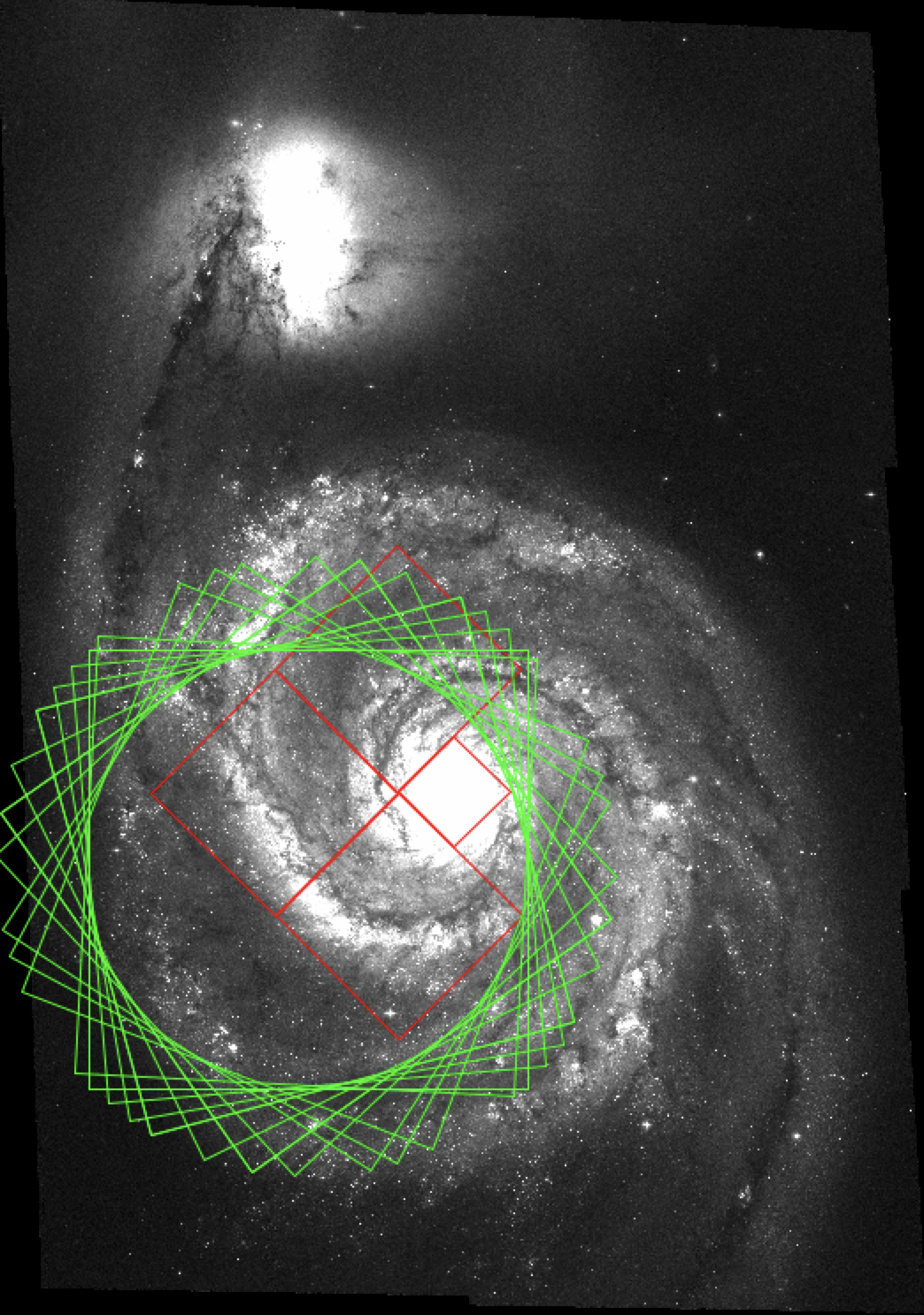}
\vspace{0.2cm}
\caption{Footprints of the Cycle 24 ACS data (green; showing a subset
  of the 34 visits for clarity) and the 1995 WFPC2 data (red),
  overlaid on the 2005 Hubble Heritage ACS B$_{435}-$band mosaic.}
\label{fig:fov}
\end{figure}

\begin{figure}[!t]
\center
\includegraphics[width=0.49\textwidth]{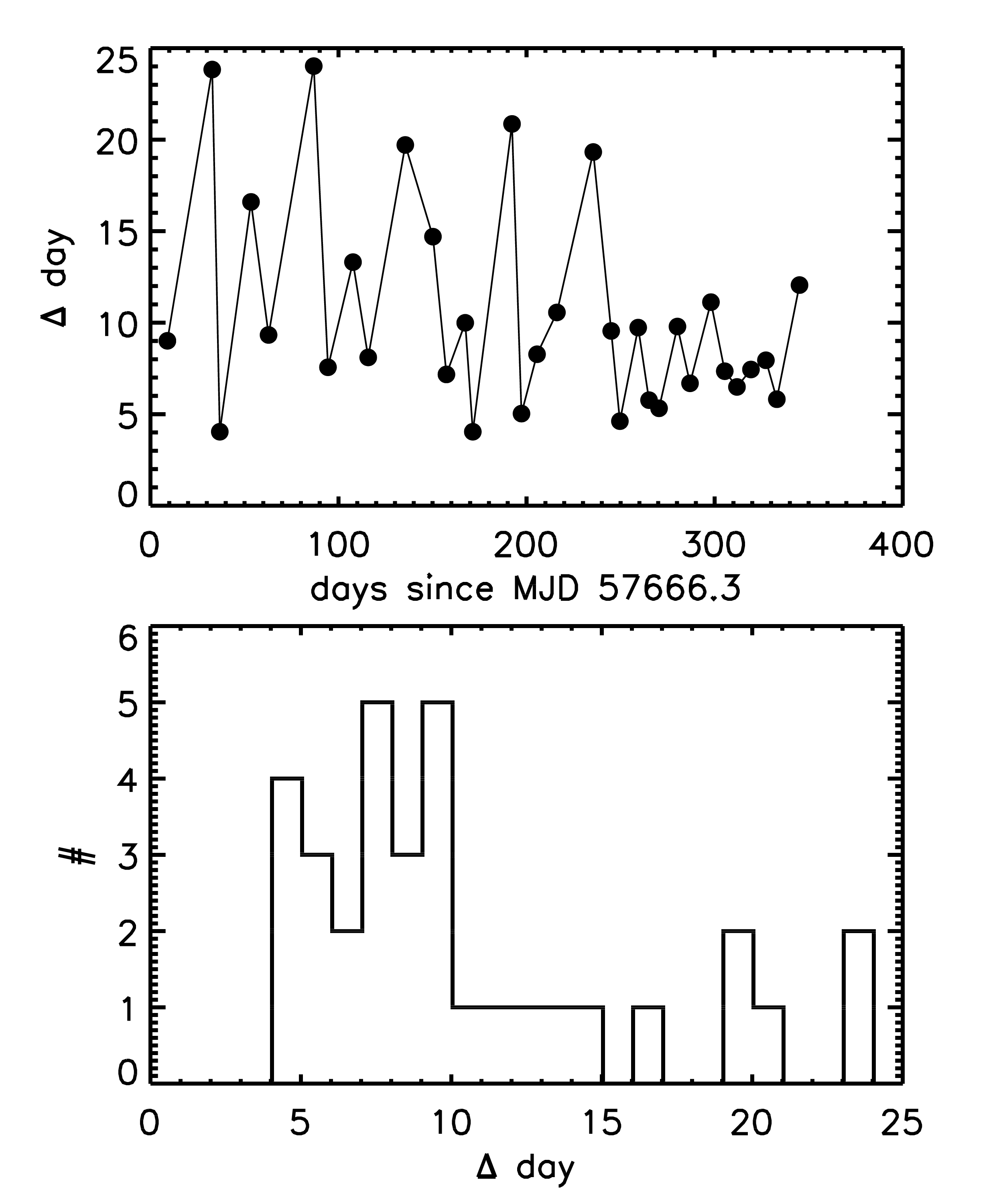}
\vspace{0.1cm}
\caption{Cadence of the Cycle 24 \HST\, data.  {\it Top Panel:}
  Separation between successive visits as a function of time.  {\it
    Bottom Panel:} Distribution of time separation between visits.}
\label{fig:cadence}
\end{figure}

\begin{figure*}[!t]
\center
\includegraphics[width=0.9\textwidth]{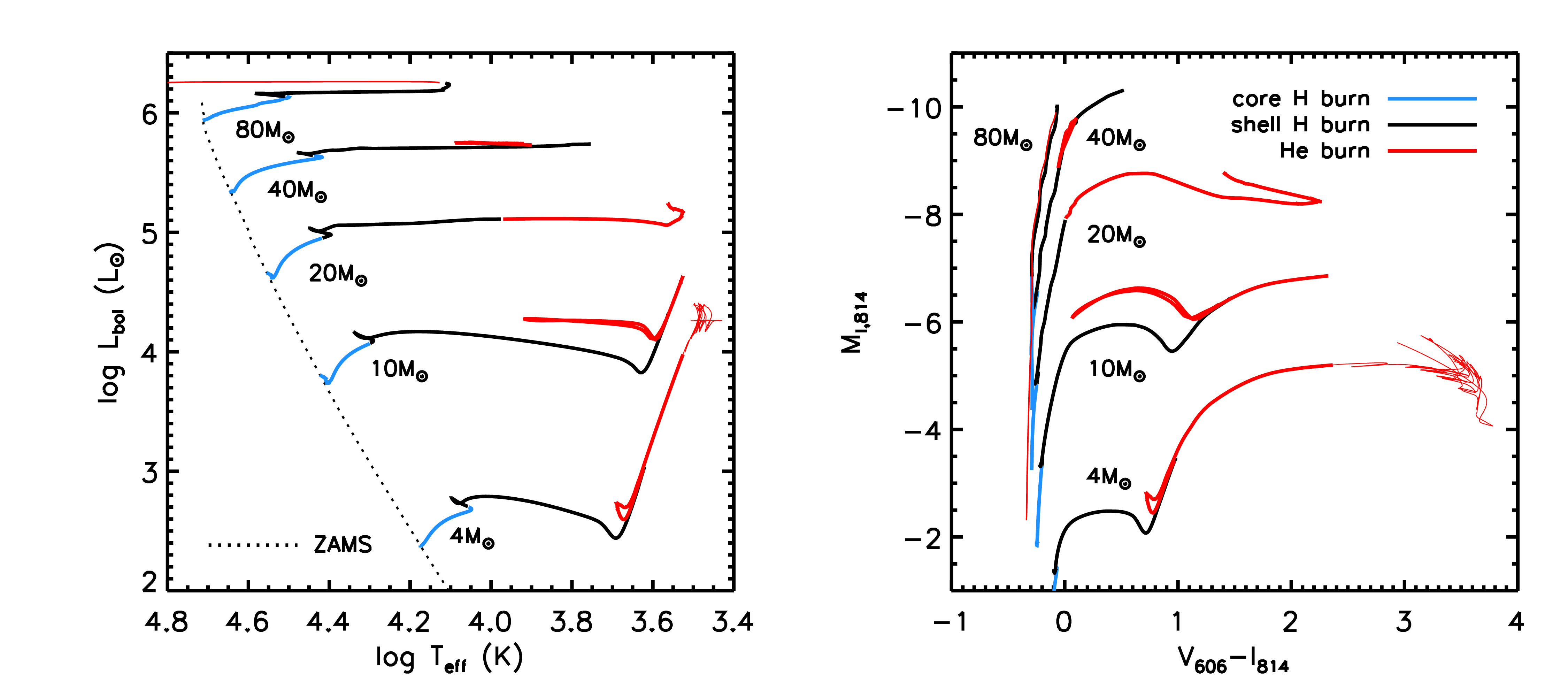}
\vspace{0.1cm}
\caption{HR diagram ({\it left}) and CMD ({\it right}) for solar
  metallicity models of 4, 10, 20, 40, and $80\Msun$ stars.
  Evolutionary phases are highlighted with different colors and
  described in the legend, and the thermally-pulsating AGB phase is
  shown as a thin line for the $4\Msun$ model.  The dotted line in the
  left panel marks the location of the ZAMS for stars from
  $1-100\Msun$.  Notice that while massive stars evolve mostly
  horizontally in the HR diagram, they brighten substantially in the
  CMD owing to large bolometric corrections for the plotted filters.}
\label{fig:mist}
\end{figure*}

From these surveys and other dedicated efforts, stellar variability
has been studied in the Galaxy, the Small and Large Magellanic Clouds,
M31 and M33, and other nearby galaxies \citep[e.g.,][]{Rejkuba03,
  Hartman06, Massey07, Massey09, Kourniotis14, Martin17, Humphreys17,
  Soraisam18}. The LMC and SMC have probably the most complete
information about stellar variability, given their close distance, low
reddening, and the long time coverage of the ongoing OGLE survey.
 
However, one of the shortcomings of studying variability in the
Magellanic Clouds, and in other dwarf galaxies such as M33, is that
they do not offer a complete view of the variability among the massive
star population. This is due to the simple fact that such stars are
rare in galaxies with low stellar masses and low star formation rates.
For example, the variability study of M33 from \citet{Hartman06} is
limited to $i>18$ ($M_i>-6.5$), and the LMC catalog of Mira variables
from \citet{Soszynski09} runs out of stars by $I\approx12$
($M_{\rm I,814}>-6.5$).

The study of more massive galaxies with higher star-formation rates
offers a more complete view of stellar variability among the most
luminous, massive stars.  The Milky Way itself is not ideal, since (i)
massive stars lie in the plane and are therefore subject to high
extinction, (ii) distances to luminous stars are not well-known, and
(iii) it is challenging to monitor the entire sky.  The {\it Gaia}
satellite will make substantial progress in addressing (ii) and (iii),
but extinction will still make it difficult to perform a complete
census, and {\it Gaia} parallaxes become quite uncertain beyond
$\approx 10$ kpc.
 
M31 mitigates some of these issues, and data from PTF have been used
to study specific classes of variables and transients. For example,
\citet{Soraisam18} performed a census of pulsating red supergiants in
M31, and \citet{Kasliwal11} presented the discovery of several faint,
fast-declining novae in M31 that could have been missed in previous
nova surveys.

Here we complement these previous studies with an investigation of
stellar variability amongst the luminous star population in M51 (the
Whirlpool Galaxy, NGC 5194), an $L^\ast$ galaxy with a stellar mass of
$\approx5\times10^{10}\Msun$ \citep{MentuchCooper12} and a star
formation rate (SFR) of $3.4\Msun$ yr$^{-1}$ \citep{Calzetti05} at a
distance of $8.6\pm0.1$ Mpc \citep{McQuinn16}.  In addition to being a
galaxy with a large population of massive stars, studying variability
within M51 offers several other benefits: the stars are all at a
common distance; the galaxy subtends a small angle on the sky,
minimizing foreground contamination; Galactic extinction is low and
fairly constant across the field; the galaxy is face-on, minimizing
internal reddening.

Perhaps the most unique aspect of this work is that our stellar
variability census of M51 is based entirely on data acquired by the
{\it Hubble Space Telescope (HST)}.  The benefits of \HST\, are many,
including a relatively well-characterized, stable PSF compared to
ground-based observations. The angular resolution (FWHM) is
$0.08-0.09\arcsec$ (depending on the filter), corresponding to
$\approx3-4$ pc at the distance of M51 and is comparable in spatial
resolution to ground-based photometry of M31.

The rest of this paper is organized as follows.  In Section
\ref{s:data} we describe the data and methods.  Section \ref{s:res}
contains our results including a census of stellar variability on
day-to-decade timescales.  We conclude with a summary in Section
\ref{s:sum}.  All magnitudes are on the Vega zeropoint system, and
absolute magnitudes are computed assuming a true distance modulus of
29.67 and Galactic extinction of $A_{\rm 606} = 0.086$ and
$A_{\rm 814} = 0.053$.  In all cases we quote $V$ and $I$ magnitudes
in the HST ACS F606W and F814W filters and refer to them as $\V$ and
$\I$, respectively.  In cases where data were originally obtained or
tabulated in another filter (such as F555W or Bessell V) we have
converted them to F606W.


\section{Data \& Methods}
\label{s:data}

\subsection{\HST\, Imaging}

The main data presented in this paper are our Cycle 24 observations,
collected in 34 epochs between Oct 2016 and Sept 2017 (Proposal ID
14704, PI Conroy).  The primary purpose of these data is to measure
the star formation history (SFH) of M51 via analysis of pixel light
curves, a technique described in \citet{Conroy15a}.  The program
required monitoring of M51 over a year, with a cadence determined by a
desire to retain sensitivity to a wide range of periods (modulated by
observatory constraints).  The resulting cadence, shown in Figure
\ref{fig:cadence}, is pseudo-random with minimum/maximum separation
between visits of 4/24 days.  Data obtained in this way offer
sensitivity to a wider range of periods than uniform sampling
\citep[see e.g.,][, and Section \ref{s:quant}]{Freedman94}.

To monitor the same field over a full year required a variation of
roll angle across the cycle (see Figure \ref{fig:fov}). We nonetheless
required that each orientation have at least two visits, to ensure
that the WFC3 parallel fields on the stellar halo would achieve
sufficient depth to detect old red giants.  The Cycle 24 data were
obtained using the Advanced Camera for Surveys (ACS) with a standard
four-point dither pattern, for a total of 2200 s per visit in each of
the $F606W$ and $F814W$.  Only the ACS data are discussed in this
paper; the parallel WFC3 data will be presented elsewhere.

To enable sensitivity to variability on timescales longer than 1 yr,
we supplement these new data with older \HST\, observations. The most
important of these are the Cycle 13 data collected in January 2005
(Proposal ID 10452, PI Beckwith) as part of the Hubble Heritage
project to obtain a six-pointing mosaic of the M51 system. A range of
filters were obtained, but here we only use the $F555W$ and $F814W$
images, collected with $4\times340$ s individual exposures in each
filter per pointing.

Finally, we also analyze Cycle 4 data obtained in January 1995
(Proposal ID 5777, PI Kirshner) to follow up SN 1994I. These data were
collected with WFPC2 in the $F555W$ and $F814W$ filters, with a single
600 s exposure per filter.  Since only a single exposure was available,
cosmic rays were removed via the L.A. Cosmic program
\citep{vanDokkum01}.

We note that many other observations have been made of M51 over the
lifetime of $HST$; we do not use these other data in this paper,
typically because they only partially overlap with our primary fields,
or are not taken in $V$ or $I$-equivalent filters.

The footprints of all the data used are shown in Figure \ref{fig:fov}.
Our Cycle 24 data contains 40\% of the total $\I-$band flux of M51 and
the WFPC2 Cycle 4 data contains 60\% of the $\I-$band flux contained in
our Cycle 24 data.

\subsection{HRDs vs. CMDs}
\label{s:cmds}

First we discuss the stars and the evolutionary phases that our
program should be able to probe.  In Figure \ref{fig:mist} we show
MIST \citep{Choi16} stellar evolutionary tracks for 4, 10, 20, 40, and
$80\Msun$ starting at the zero-age main sequence (ZAMS, shown as a
dotted line).  The models are at solar metallicity.  The tracks are
color-coded by their evolutionary phase.  The left panel shows the
behavior in the HR diagram while the right panel shows the same models
in the VI color-magnitude diagram (CMD).  

The bolometric corrections for massive stars are substantial,
especially for the hot stars.  For example, a $20\Msun$ star evolves
at approximately constant $L_{\rm bol}$ but brightens in
$M_{\rm I,814}$ by $>5$ mag.  Moreover, stars more massive than
$\approx40\Msun$ do not at any point exceed $M_{\rm I,814}\approx-10$
and indeed during the main sequence they are relatively faint in $\I$:
an $80\Msun$ star has $M_{\rm I,814}\approx-5.5$ at the ZAMS.  This
leads us to an important point that will be relevant for interpreting
the results in later sections: essentially every star brighter than
$M_{\rm I,814}\approx-6$ is an {\it evolved} high-mass star.

\subsection{Photometry}

\subsubsection{PSF photometry with DOLPHOT}
\label{s:psfvsap}

\begin{figure}[!t]
\center
\includegraphics[width=0.49\textwidth]{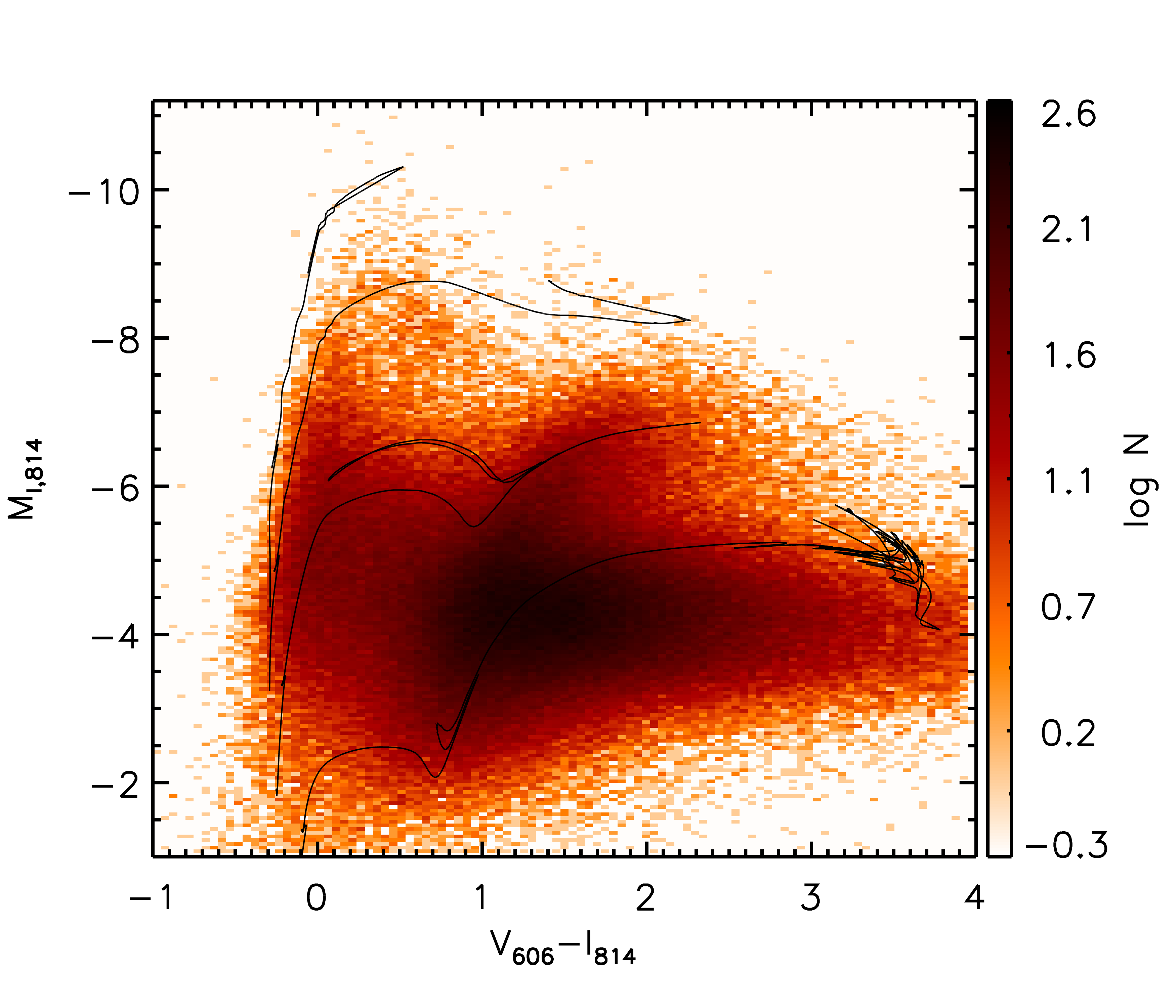}
\vspace{0.1cm}
\caption{CMD of the entire bright star population in M51 from our
  Cycle 24 data.  Model tracks at 4, 10, 20, and $40\Msun$ are shown
  for reference.}
\label{fig:basic}
\end{figure}

\begin{figure}[!t]
\center
\includegraphics[width=0.49\textwidth]{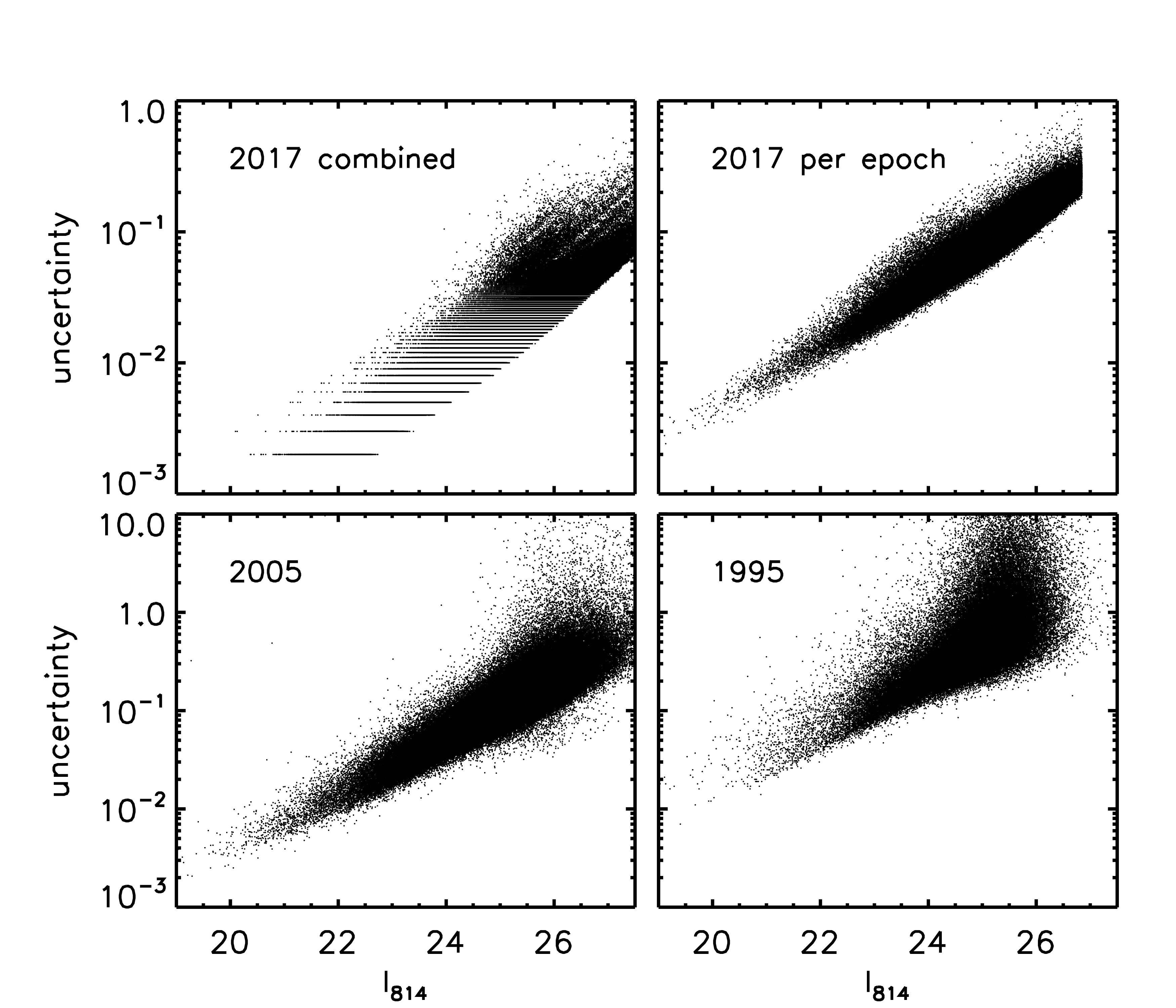}
\vspace{0.1cm}
\caption{Photometry uncertainty reported by DOLPHOT as a function of
  $\I-$band magnitude.  Upper panels show our Cycle 24 ACS data,
  combined over all 34 epochs (upper left) and the median per epoch
  (upper right).  Lower panels show archival ACS imaging in 2005
  (lower left) and WFPC2 imaging in 1995 (lower right).}
\label{fig:errors}
\end{figure}

We perform PSF photometry using the DOLPHOT software package
\citep{Dolphin00}.  Briefly, DOLPHOT performs photometry by
iteratively identifying stars, and simultaneously solves for the sky
background and the magnitude of each star for an arbitrary number of
aligned images in an arbitrary number of filters.  DOLPHOT uses the
\texttt{TinyTim} PSF models \citep{Krist95} and makes additional
adjustments to the PSF using relatively bright, isolated stars.
DOLPHOT also computes and applies aperture corrections.  We supply to
DOLPHOT CTE-corrected individual exposure-level images for the ACS
data (\texttt{flc} files) and standard exposure-level images for the
WFPC2 data (\texttt{c0m} files).

There are many parameters to set when running DOLPHOT.  We adopt the
parameters used for PSF photometry of M31 stars in the \HST\, PHAT
Survey \citep{Dalcanton12, Williams14}.  We refer the interested
reader to those papers, which describe extensive tests of the
impact of various DOLPHOT parameters on the derived photometry.

After experimentation with a variety of analysis procedures, we
decided to process the images in two DOLPHOT runs.  The first contains
the Cycle 24 data in both filters, and the second contains the Cycle
4, Cycle 13, and four visits from the Cycle 24 data to assist with the
subsequent cross-matching between the two runs.  This process ensured
that stars that were well-detected in certain epochs/filters would
have measured photometry in all available epochs/filters even in cases
where the individual exposure S/N was low.  The runs were split in two
for computational efficiency (the Cycle 24 photometry took three weeks
to complete).  The catalogs were matched in pixel coordinates and we
require an association of better than one pixel for a match, although
in nearly all cases the match is within $<0.3$ pixel.  DOLPHOT
measures and applies aperture corrections such that the reported
photometry is within a radius of $0.5\arcsec$.  We apply additional
aperture corrections such that the final photometry is reported within
an infinite aperture \citep[see][]{Bohlin16}.

DOLPHOT provides a number of parameters for each star, including a
statistic of the goodness of fit ($\chi$), several shape parameters
(round and sharp), an ``object type", and a measure of the crowding.
This last parameter is defined in magnitudes and quantifies how much
brighter the star would be if nearby stars had not been fit
simultaneously.  Unsurprisingly, the crowding is strongly correlated
between the $\V$ and $\I$ bands with a ($3\sigma$ clipped) standard
deviation of 0.15 mag.  We require an object type$=1$ (star),
$\I-$band crowding $<0.5$, $\I-$band sharpness, $-0.2<$sharp$<0.2$,
and $\chi_I<2$.  We also require a final $\I-$band S/N$>5$ and
detections in the $\I-$band for least five epochs. We have also masked
the central $10\arcsec$ and regions around six bright foreground
stars.

With these cuts our final catalog contains 320,000 stars, of which
73,000 are brighter than $M_{\rm I,814}=-5$.  98\% of stars brighter
than $I=26$ have an identified $\V-$band counterpart in the Cycle 24
data.  The cross-matches between Cycle 24 $I$-band and the archival
$I-$band in the regions of overlap drop below 80\% for $I>25.5$ and
$I>23.5$ for the 2005 and 1995 data, respectively.  This is mostly a
reflection of the shorter exposure times in the archival data.

We comment briefly on possible confusion between star clusters and
bright stars.  The angular resolution (FWHM) of the $\V$ ($\I$) data
is $0.08\arcsec$ ($0.09\arcsec$), which at the distance of M51
corresponds to 3.33 (3.75) pc.  For high S/N sources it is possible to
separate stars from star clusters to a fraction of a resolution
element.  Based on \HST\, analysis of star clusters in M31,
\citet{JohnsonC12} find sizes of the brighter clusters in the range of
$1-10$ pc with very few below 1 pc. Our sample of stars is therefore
unlikely to be contaminated by unresolved star clusters.  However,
unresolved similar-mass binaries will be a source of contamination in
our catalog; such sources can only be identified with followup
spectroscopic monitoring.

Corrections for Galactic extinction were applied to all photometry
adopting the \citet{Schlafly11} reddening map.  Specifically, we adopt
$A_{\rm 555} = 0.097$, $A_{\rm 606} = 0.086$,
$A_{\rm 814} = 0.053$.  We convert all of the $V_{\rm 555}$
photometry to $\V$ using $V_{\rm 555}-\I$ and the bolometric
correction tables distributed as part of the MIST project
\citep{Choi16}.

An overview CMD derived from the Cycle 24 data is shown in Figure
\ref{fig:basic}.  One clearly sees the locus of main sequence and blue
core He burning stars at $\V-\I\approx0$, the red core He burning
stars (RSGs) occupying the diagonal sequence at $1<\V-\I<2$ and
$-7<M_{\rm I,814}<-4$, and the asymptotic giant branch (AGB) stars at
$\V-\I>2$.  incompleteness starts to become substantial at
$M_{\rm I,814}>-4$, especially for the redder stars.

\subsubsection{Error budget}

The error budget is critical for an accurate accounting of stellar
variability.  In Figure \ref{fig:errors} we show the reported errors
from DOLPHOT for the 1995, 2005, and 2017 (Cycle 24) data.  For the
latter we show two versions: the median (over the 34 visits) per epoch
error, and the error on the final combined photometry.  The
uncertainties depend on several factors, including the photon noise,
sky background level, and goodness of fit of the PSF model.  Notice
that the errors continue to decline toward brighter magnitudes.

\begin{figure}[!t]
\center
\includegraphics[width=0.49\textwidth]{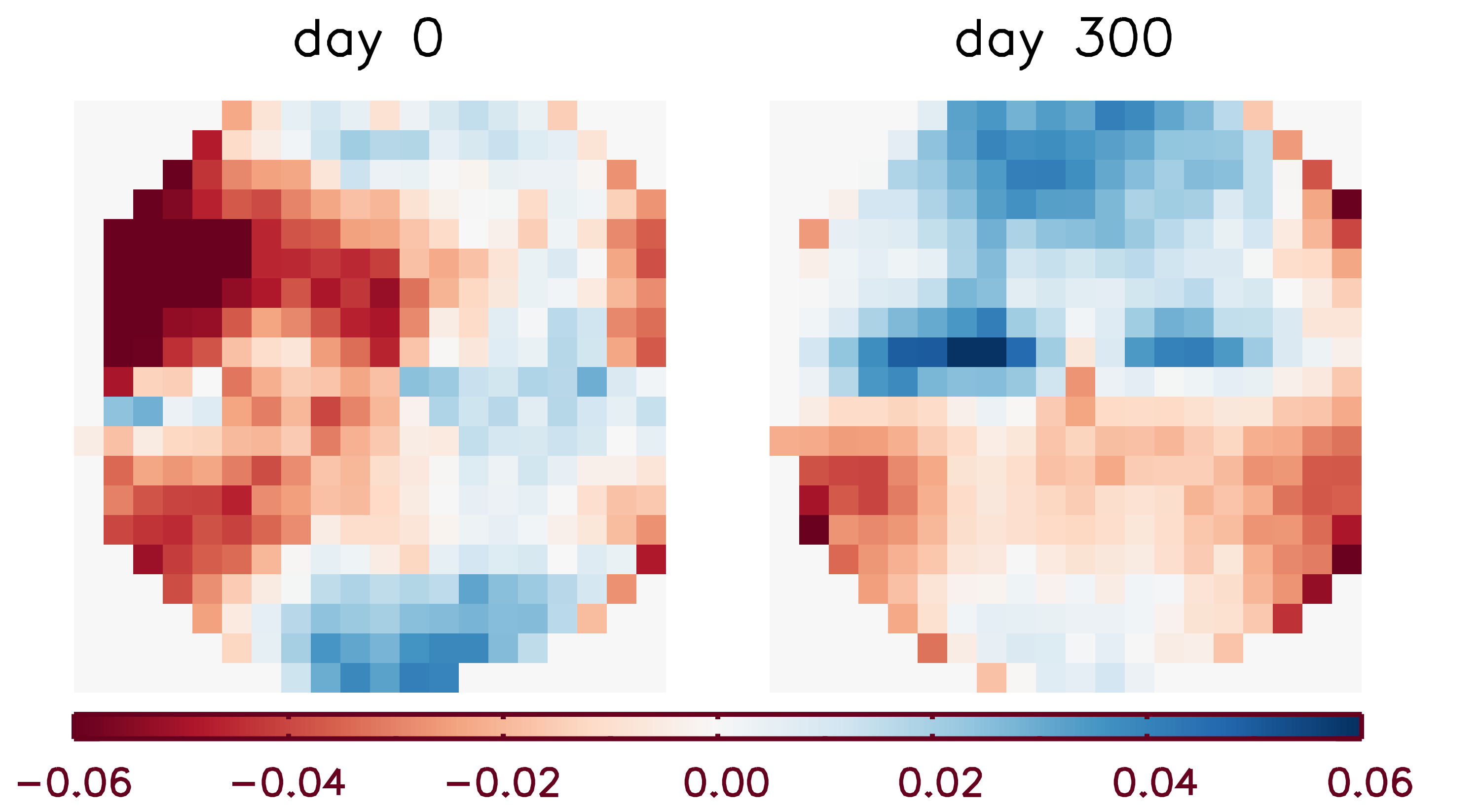}
\vspace{0.1cm}
\caption{Maps of mean magnitude corrections in the $\I-$band as a
  function of position in the ACS image frame at two epochs (indicated
  as days since the first visit).  Stars from all 34 visits were
  grouped in $200\times200$ pixel regions in the image plane and mean
  $\Delta m$ values were computed as a function of visit MJD.  For
  each star, $\Delta m$ is the difference between its mean magnitude
  across all 34 visits and the magnitude at that visit.  Quantities
  were computed for pixel regions containing $>1000$ stars.  Notice
  the strong time and spatial dependence of the mean magnitude
  corrections.  At day 300 the magnitude correction is discontinuous
  across the chip gap (which runs horizontally through the center of
  these maps).}
\label{fig:ucal}
\end{figure}

As we will see in later sections, many bright stars have rms
variations in their light curves at the $\approx0.01-0.1$ mag level,
so understanding additional sources of uncertainty not included in the
default DOLPHOT output is important.  To our knowledge all
sub-percent-level variable star photometry has relied on differential
photometry that employs non-variable stars as calibrators
\citep[e.g.,][]{Hartman04, Nascimbeni14, Jayasinghe18}.  In our case
this is not possible since we have reason to believe that the majority
of bright stars are intrinsically variable.  We therefore undertake a
variety of tests of the absolute stability of the DOLPHOT PSF
photometry, noting that previous work has found evidence for
systematic uncertainties at the $\approx0.03$ mag level
\citep[e.g.,][]{Dalcanton12, Williams14}.

To test the DOLPHOT photometry we have selected a sample of bright,
uncrowded stars with $\I<22.6$ ($M_{\rm I,814}<-7$) and an $\I-$band
crowding parameter $<0.02$.  We ran many permutations of the default
DOLPHOT parameters including varying the sky subtraction model, the
radius used to fit the PSF and the sky background, turning on/off the
aperture corrections, and considering aperture photometry instead of
PSF photometry.  Nearly all of the changes had no discernable effect
on the light curve rms, with the exception of aperture photometry with
an aperture radius of 3 pixels.  This choice of parameters resulted in
a lower rms on average for $0.02<\Irms<0.05$.  Larger and smaller
values of $\Irms$ did not change.  This test suggests that the PSF
model is insufficiently flexible to account for subtle changes in
either time or location on the detector \citep[see
also][]{Dalcanton12, Williams14}.  Unfortunately we cannot use
aperture photometry for the bulk of our analysis because the vast
majority of stars are in very crowded regions where simultaneous
fitting of overlapping PSFs is essential for accurate photometry.

The large number of distinct orientations (17) allows us to perform a
novel test of the reliability of the PSF photometry without relying on
calibration from non-variable stars.  If we consider a small region in
the image plane, then as the orientation varies the stars at that
position will also vary.  Over the course of the full Cycle there will
be 17 distinct populations of stars landing within a given image plane
region.  If the photometry were free of systematic errors, then the
mean magnitude offset of the stars at each visit with respect to their
individual mean magnitudes should be zero, i.e.,
$\sum_i(m_{i,j}-\bar{m_i})/N_j$ should be zero where $m_{i,j}$ is the
magnitude of the $i-$th star at the $j-$th orientation, $\bar{m_i}$ is
the mean magnitude of star $i$ across all visits, and $N_j$ is the
number of stars at the $j-$th orientation.  The sum is over all stars
within a given small region in the image plane.

\begin{figure}[!t]
\center
\includegraphics[width=0.49\textwidth]{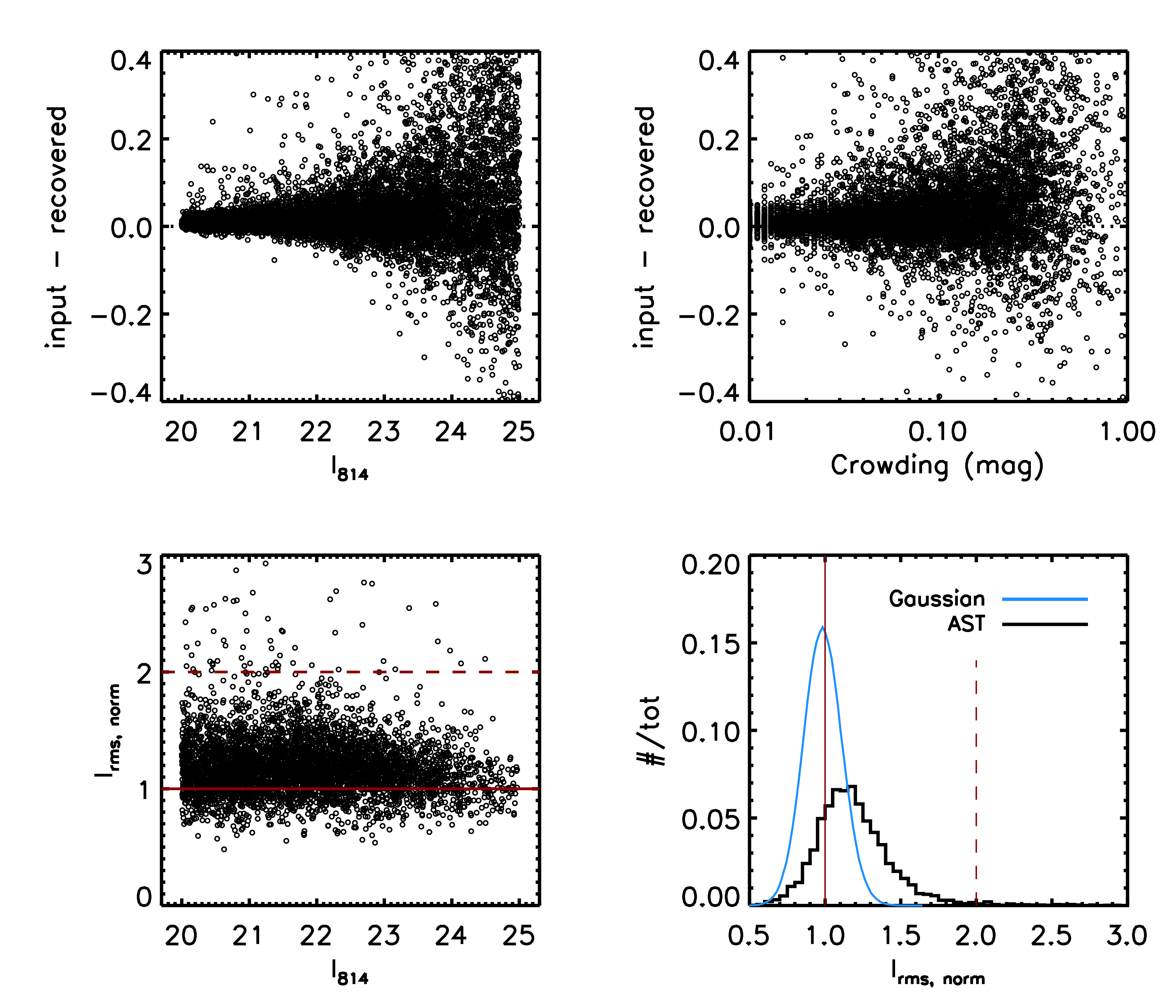}
\vspace{0.1cm}
\caption{Artificial star tests.  {\it Upper Panels:} Comparison of
  input and recovered $\I-$band magnitudes as a function of input
  magnitude ({\it left}) and crowding ({\it right}). The crowding
  parameter measures how much brighter a star would be if surrounding
  stars were not simultaneously fit.  {\it Bottom Left:} rms of the
  error-normalized light curve for each artificial star, shown as a
  function of input magnitude.  Here only stars with crowding$<0.1$
  are shown.  The dashed line is our threshold for variability.  {\it
    Bottom Right:} Histogram of the rms of the error-normalized light
  curves for the artificial stars (black line) compared to draws from
  a unit Gaussian (blue line).  If the reported errors accurately
  reflect the total error budget and if the errors are Gaussian then
  the solid and dashed lines should be equal.}
\label{fig:fake}
\end{figure}

In practice we find that these mean magnitude offsets, or magnitude
corrections, vary with time and with position in the image plane.  We
have computed magnitude correction maps in $200\times200$ pixel
regions across the image plane in both the $\V-$ and $\I-$band.  We
then fit quadratic functions to the time-dependence of the corrections
for regions containing $>1000$ stars (combined over all visits).
These maps are shown in Figure \ref{fig:ucal} for the $\I-$band at two
epochs.  The corrections range from $-0.06$ to $+0.06$ and vary in a
systematic and generally smooth manner, although there are often sharp
transitions across the chip gap (which runs horizontally through the
center of these maps).  We have applied these corrections to all our
photometry and find that for $0.02<\Irms<0.1$ these corrections result
on average in smaller $\Irms$ values.  For values outside this range
there is little change.  As a final measure we add a systematic error
of 0.02 mag to the quoted DOLPHOT uncertainties to capture residual
issues at this low level.

\begin{figure}[!t]
\center
\includegraphics[width=0.49\textwidth]{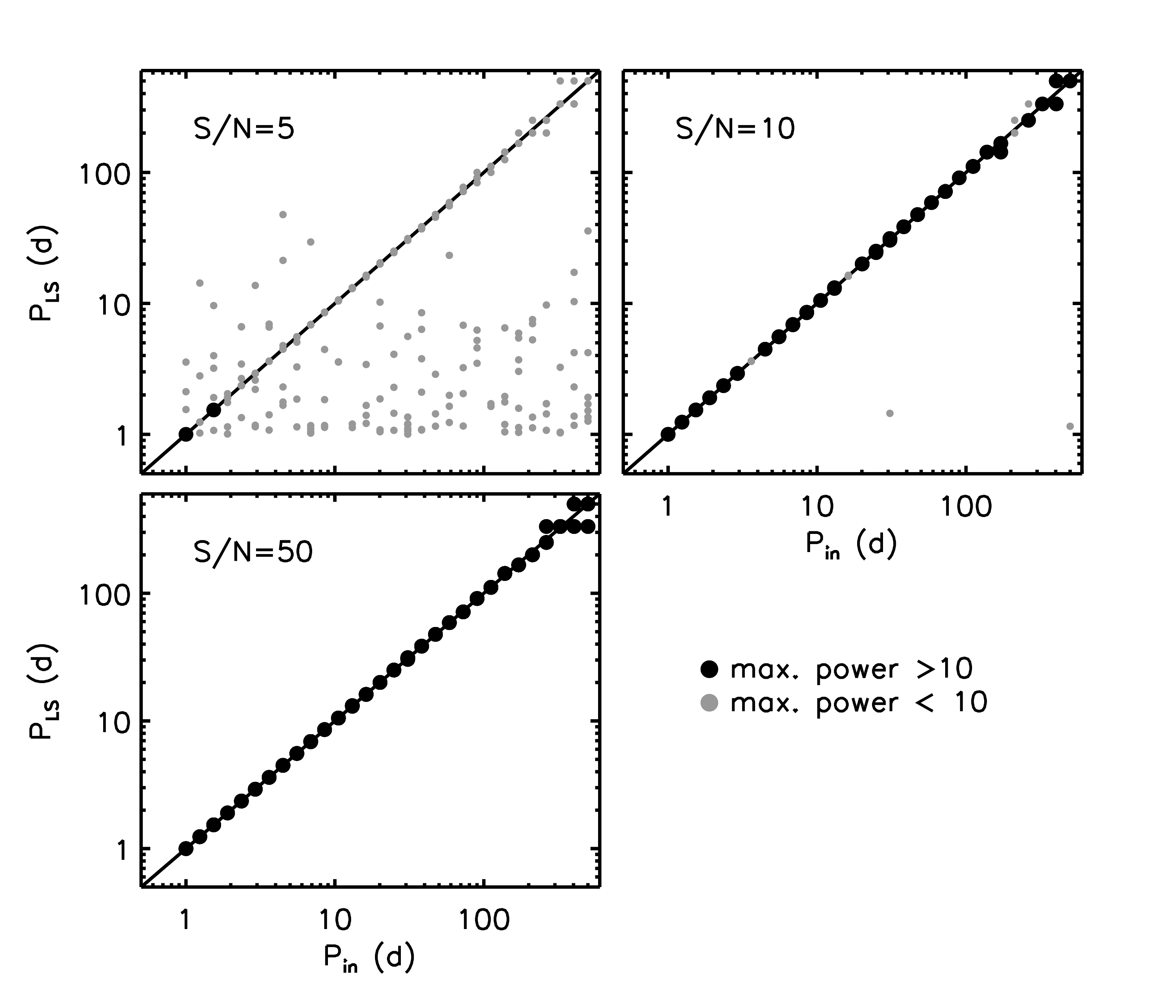}
\vspace{0.1cm}
\caption{Recovery of periods based on Lomb-Scargle periodogram
  analysis.  Mock data were generated from sine waves with the same
  cadence as our Cycle 24 data.  Each panel shows the recovery for a
  given S/N per epoch.  Each input period was simulated for 20 phases
  randomly drawn between $0-2\pi$.  Black (grey) points show results
  where the maximum power in the periodogram is $>10$ ($<10$).}
\label{fig:lssim}
\end{figure}

\begin{figure*}[!t]
\center
\includegraphics[width=0.95\textwidth]{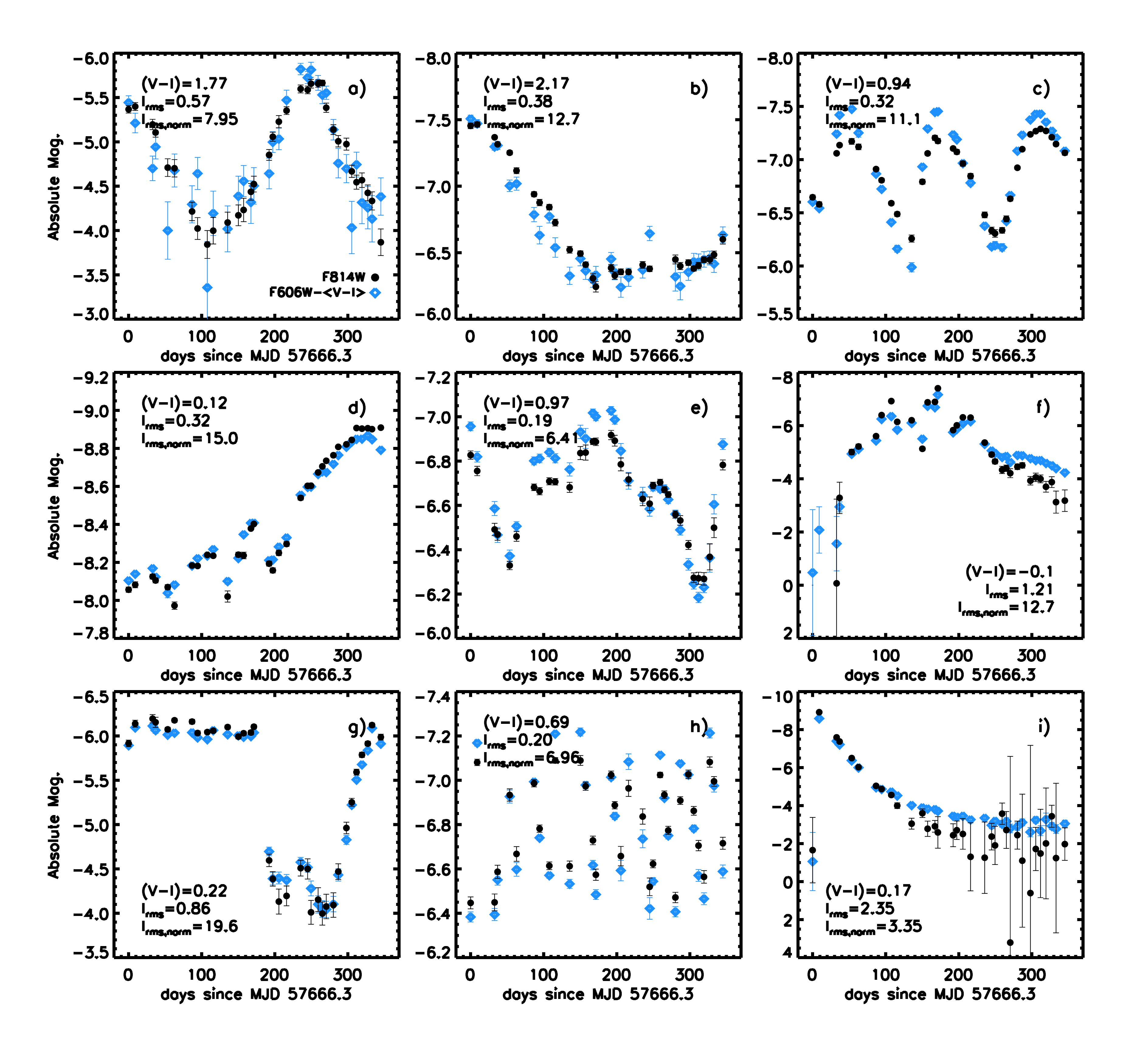}
\vspace{0.1cm}
\caption{Selected light curves displaying a variety of interesting
  behavior. Both $\I-$band $\V-$band magnitudes are shown (black and
  blue symbols), where the latter has been shifted by the mean $\V-\I$
  color for clarity.  Also shown in the legend are the $\I-$band rms of
  the light curve, $\Irms$, and the mean $\V-\I$ color. Panel a) and b)
  typical behavior of luminous red stars; c) an RV Tau star; d) and e)
  large amplitude luminous blue star variability; f) an extreme blue
  variable; g) an R Cor Bor star; h) a bright Cepheid; i) an extremely
  luminous nova.  Stars are sorted by $\Irms$.  See the Appendix for
  additional example light curves.}
\label{fig:blue}
\end{figure*}

The apparent time variation in the magnitude corrections may be due to
some subtle underlying systematic, such as residual CTE effects
combined with unmodeled variation in the point spread function. With
very careful cosmic ray removal and large-aperture photometry of
individual bright stars we were able to reduce the variation to
$<0.005$ after correcting for residual correlations in the image
plane, but this methodology cannot be applied "in bulk" or for crowded
regions.  Even though we have not been able to pinpoint the source of
the variation, we stress that a) these same patterns are seen in
previous studies, in particular PHAT \citep{Dalcanton12} and b) we can
correct for them even if we do not understand the source.  We also
note that our main conclusions are not sensitive to these corrections,
although they do affect the fractions of variable stars in the bin of
lowest variability ($\Irms\approx0.03$ mag).

A common technique for assessing the quality of the DOLPHOT photometry
is the use of artificial star tests \citep[e.g.,][]{Dalcanton12,
  Williams14}.  The general idea is to inject stars with known fluxes
into the image and to photometer those stars as usual with DOLPHOT to
test how well the fluxes can be recovered as a function of magnitude
and local stellar density.  We stress that the injection of artificial
stars only tests some of the possible sources of systematic
uncertainties.  For example, artificial stars are injected with the
same model PSF used in the fitting, so this test cannot assess the
systematic uncertainty associated with possible PSF mis-match.

We injected $10^4$ fake stars with a uniform distribution of
magnitudes from $I_{\rm 814}=20-25$ and sampling a range of local
stellar densities.  The results are shown in Figure \ref{fig:fake}.  In
the upper panels we compare the input and recovered magnitudes and
do not identify any significant biases as a function of magnitude but
do find a bias as a function of the crowding parameter such that more
highly crowded regions result in slightly brighter recovered fluxes
compared to the input.

The lower panels quantify the variability of the fake star light
curves.  Recall that the injected stars do not have any intrinsic
variability.  We define a quantity, $\Inorm$ as the rms of the
error-normalized light curve, i.e., rms$(m_j/e_j)$ where $m_j$
and $e_j$ are the magnitudes and errors of the $j-$th visit in the
light curve.  This quantity is plotted as a function of input
magnitude in the lower left panel and as a histogram in the lower
right panel.  Here we focus only on stars with crowding$<0.1$.  In the
right panel we also show the expected distribution if the reported
errors captured the entire error budget by drawing mock light curves
with the reported errors and measuring the normalized rms values from
those mock data.  As expected, this distribution is centered on 1.0
and has a width determined by the finite number of samples from the
light curve (34).

The lower right panel motivates us to consider a value of
$\Inorm=2.0$ as a threshold above which the light curve variation
reported by DOLPHOT is likely to be real.  In our tests $<2$\% of
artificial stars have $\Inorm>2.0$. This threshold is also plotted in
the lower left panel.

In summary, we have applied magnitude corrections to the Cycle 24
photometry that are a function of detector position and time.  In
addition, we have added 0.02 mag in quadrature to the DOLPHOT reported
errors, to capture a host of possible systematic uncertainties at this
low level.  With these modifications we expect the vast majority of
stars with $\Inorm>2.0$ to display genuine temporal variability.
Note that for the brightest stars the error budget is dominated by the
0.02 mag systematic uncertainty, so $\Inorm=2.0$ corresponds to a
threshold of $\Irms=0.035$.

\begin{figure*}[!t]
\center
\includegraphics[width=0.95\textwidth]{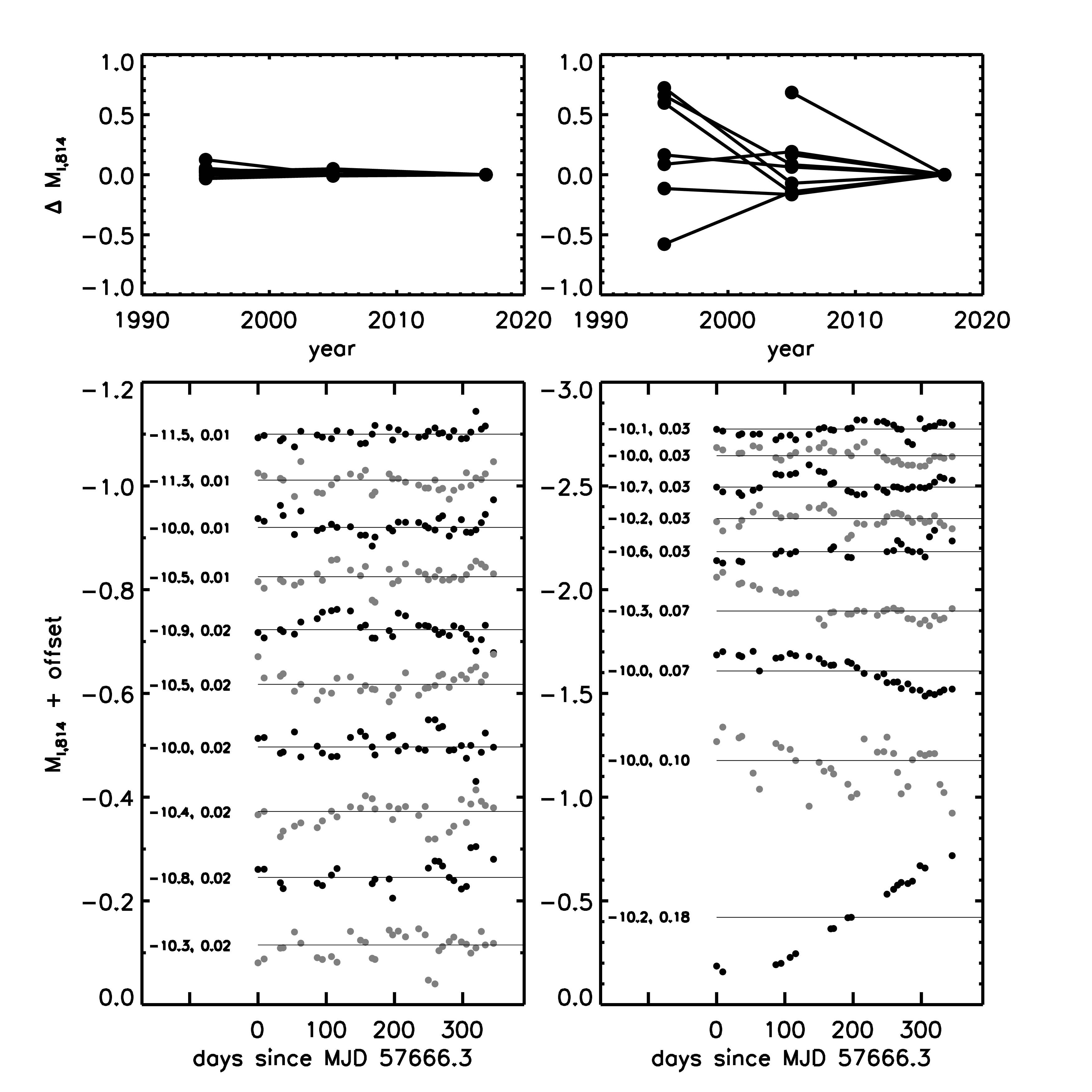}
\vspace{0.1cm}
\caption{$\I-$band light curves of the 19 brightest stars in M51
  ($M_{\rm I,814}<-10$).  The bottom panels show the Cycle 24 data
  sorted by $\Irms$.  Each light curve is offset for clarity and is
  labeled by $M_{\rm I,814}$ and $\Irms$.  The upper panels show the
  change in magnitude between 1995, 2005, and 2017 for the same light
  curves shown in the bottom panels.  Here the light curves are
  normalized to the 2017 data.  Notice the different y-axis ranges in
  the lower left and right panels, though in both panels the y-axis
  tick marks are spaced at 0.1 mag intervals.  Stars in the right
  panels show clear evidence for variability on both month-year and
  decade timescales.  Amongst these brightest stars the variability
  fraction is therefore 9/19$\approx50$\%.}
\label{fig:bright}
\end{figure*}

\begin{figure*}[!t]
\center
\includegraphics[width=0.95\textwidth]{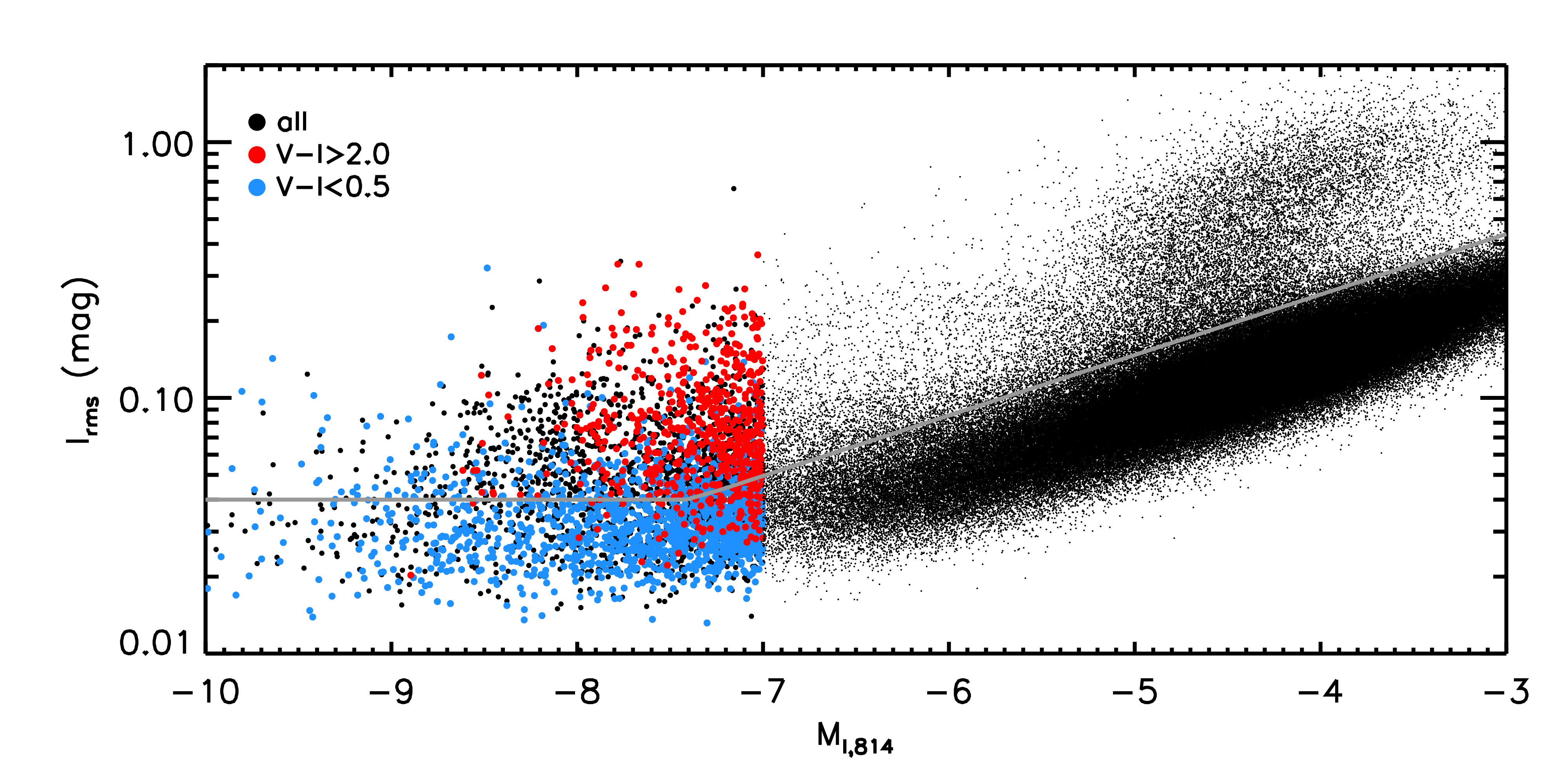}
\caption{$\I-$band light curve rms as a function of $\I-$band
  magnitude for 240,000 stars selected to have at least 20 epochs of
  data and a crowding parameter $\leq 0.5$.  Also shown in grey is an
  approximation of the threshold used to define variability
  ($2.0\times$ the median per-epoch error).  The plume of stars at
  high $\Irms$ and $M_{\rm I,814}>-5.5$ are Mira and SRV variables.
  Bright stars are color-coded by $\V-\I$ color ($>2.0$, red symbols;
  $<0.5$, blue symbols). }
\label{fig:vfrac1}
\end{figure*}

\begin{figure}
\center
\includegraphics[width=0.45\textwidth]{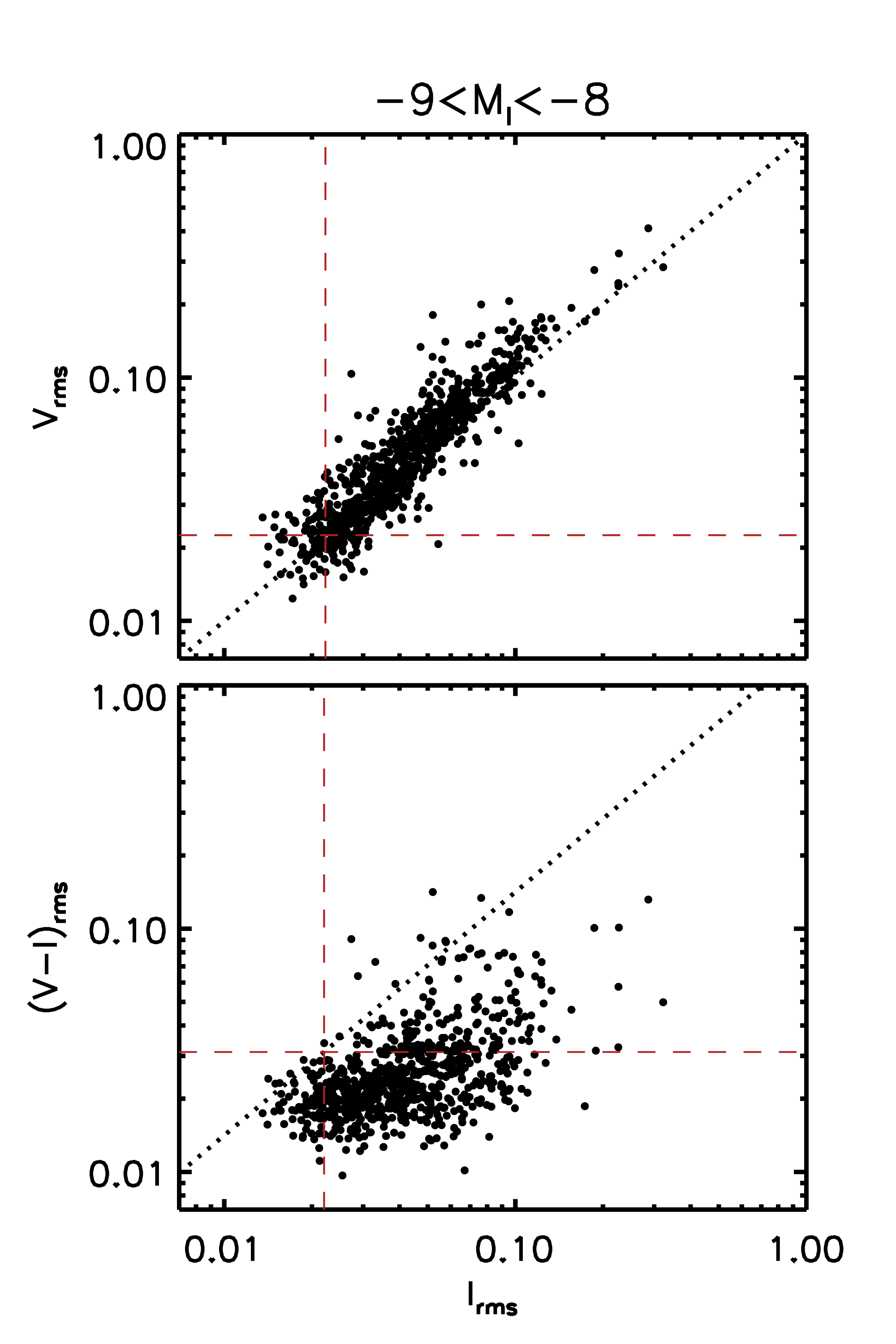}
\caption{{\it Top Panel:} Comparison between $\Irms$ and $\Vrms$ for
  stars with $-9<M_{\rm I,814}<-8$.  Dotted line is the relation $y=x$
  and dashed lines are the median errors of the sample in the $\V$ and
  $\I$ bands.  {\it Bottom Panel:} Comparison between $\Irms$ and
  $(V-I)_{\rm rms}$.  Dotted line is the relation $y=\sqrt{2}x$, and
  the dashed lines are the median errors in $\I$ and $(V-I)$.  A
  narrow magnitude range has been plotted here so that the range of
  photometric uncertainties is also narrow - variation in the
  quantities plotted is therefore driven by genuine photometric
  variability.  The very small color variation (bottom panel)
  indicates that most of the stars shown here harbor coherent
  variation in the $\V$ and $\I-$band light curves (see Appendix
  \ref{s:exlc} for examples). }
\label{fig:varvi}
\end{figure}

\begin{figure*}[!t]
\center
\includegraphics[width=0.9\textwidth]{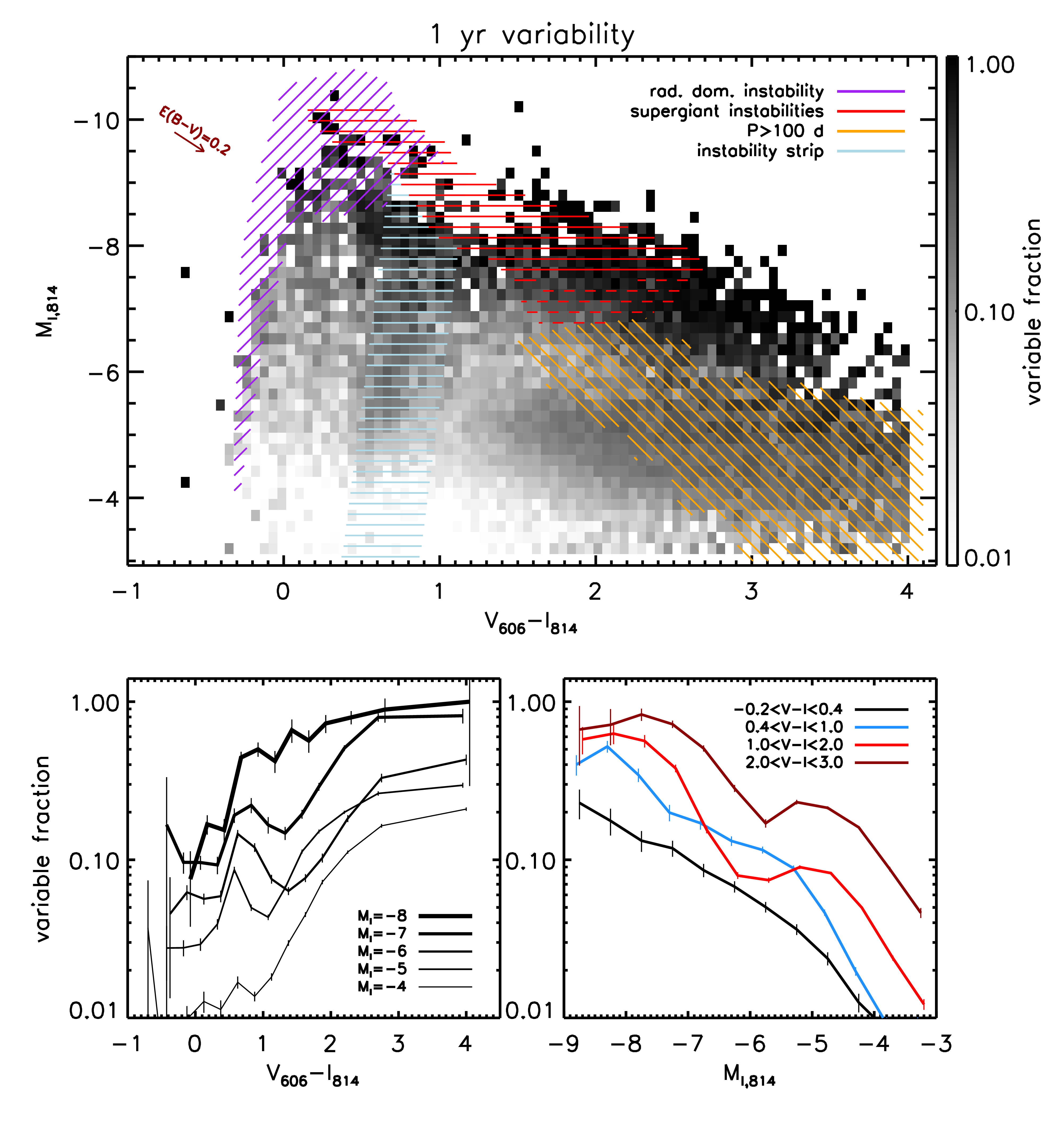}
\vspace{0.1cm}
\caption{Variability fraction across the CMD on year timescales. Here
  the variable fraction is defined conservatively as stars with
  $I_{\rm rms,norm}>3$.  {\it Top Panel}: Variable fraction in
  color-magnitude bins.  The instability strip at $\V-\I\approx0.7$ is
  clearly visible, as are the red supergiants and AGB stars at
  $\V-\I>2$ and the luminous blue star population at
  $M_{\rm I,814}<-7$. At bright magnitudes the fraction is noisy due
  to the small number of stars per bin.  Regions where variability are
  expected on theoretical grounds are indicated as hatched regions;
  see text for details.  A reddening vector for $E(B-V)=0.2$ is also
  shown.  {\it Bottom Panels}: Variable star fraction plotted in bins
  of magnitude as a function of color ({\it left panel}), and in bins
  of color as a function of magnitude ({\it right panel}).}
\label{fig:vfraccmd}
\end{figure*}

\subsection{Quantifying variability}
\label{s:quant}

There is a rich literature describing numerous techniques for
quantifying variability in astronomical light curves
\citep[e.g.,][]{Stetson96, Kim14, Jayasinghe18}.  Broadly, these
techniques fall into two categories: non-parametric statistics of the
light curves, e.g., the rms and higher order moments, and
period-finding statistics, e.g., the Lomb-Scargle periodogram
\citep{Lomb76, Scargle82, VanderPlas18}.

Here we consider both approaches to quantifying variability.  In the
first, we measure moments of the light curve, focusing primarily on
the rms.  We also consider an error-normalized version of the rms, as
defined in the previous section.  These quantities are referred to as
$\Irms$ and $\Inorm$ for the values computed from the $\I-$band light
curve.  In practice we compute the rms (and the error-normalized rms)
after iteratively $3\sigma$ clipping the data. 

The second approach we consider is applying Lomb-Scargle periodogram
analysis to the light curves.  For this purpose we use the IDL program
\texttt{scargle.pro}.  We searched for periods between $1-400$ d using
10,000 frequencies and a false-alarm probability of 0.01.  In order to
test our ability to recover periods with our cadence and S/N we
performed a set of simulations.  We created mock light curves with the
cadence of our observations at S/N$=5,10,50$ for a range of periods
and phases.  The light curves are sine waves with unit amplitudes.

The result of this test is shown in Figure \ref{fig:lssim}, where we
compare the input and recovered periods for the three S/N values.  In
each panel there are ten points at each input period representing the
ten randomly drawn phases.  Points are color-coded by whether their
maximum power is $>10$ or $<10$, which is close to a false-alarm
probability of 0.01.  It is clear that by S/N$=10$ we are able to
recover nearly all input periods to high fidelity.  Note that our
pseudo-random cadence enables sensitivity to periods at least as short
as one day (and likely shorter) for regular, sinusoidal light curves.
Furthermore, while we are able to recover periods $>100$ d even for
data that only span one year, we note that this success is due in part
to the very simple input light curve shape.  In practice, we treat the
longer periods ($>100$ d) merely as indicative of long timescale
variability in the light curve.

We close this section with a few words of caution regarding the
Lomb-Scargle analysis.  As is well-known, the interpretation of
false-alarm probabilities in periodograms is fraught with difficulty
\citep[][and reference therein]{VanderPlas18}.  The test performed in
Figure \ref{fig:lssim} provides some assurance that a threshold of
maximum power $>10$, which corresponds approximately to a false-alarm
probability of $<0.01$, is adequate for identifying sine-wave
variability for data with S/N>10.  In reality the light curves are
often not sinusoidal.  However, inspection of thousands of light
curves has lead us to conclude that this threshold provides a
satisfactory balance between minimizing false-positives and
identifying light curves with clear variability.  Nonetheless we
caution that results relying on a light curves selected by maximum
power in the periodogram (principally Figure \ref{fig:periodmag}) will
likely suffer from some degree of contamination and incompleteness.


\vspace{2cm}

\section{Results}
\label{s:res}

\subsection{Variability on day-to-year timescales}

In this section we focus on variability measured within our Cycle 24
data, which has sensitivity to variability on day-to-year timescales.

\subsubsection{Example light curves}

In Figure \ref{fig:blue} we highlight unusual and interesting light
curves identified in the data.  In each panel we show the $\V-$ and
$\I-$band data, with the former shifted by the mean $\V-\I$ color of
the star (recall that the photometry has been corrected for Galactic
extinction but not for reddening within M51).  We also list the mean
color and $\Irms$.  One clearly sees a wide range of behavior
including periodic light curves, long-term variation, and relatively
erratic changes.  Specifically, we have selected two fairly typical
luminous red giants (a,b), an RV Tau star (c), two luminous blue
variables (d-e), an extreme amplitude blue variable (f), an R Coronae
Borealis star (g), a bright Cepheid (h), and a very luminous nova (i).
R Coronae Borealis (RCB) stars are a class of rare, hydrogen-deficient
carbon rich supergiant variables.  The RCB star in panel (g), at
$M_{\rm I,814}=-6.0$ is perhaps the brightest known star of this class
\citep{Tisserand09, Tang13}.  Our dataset contains at least five novae
and half a dozen very bright RV Tau stars ($-7<M_{\rm I,814}<-6$).

In Figure \ref{fig:blue} the color variation is small, and in many
cases almost absent, even in the presence of large amplitude
variations.  For the red stars this suggests that we are witnessing
primarily variation in the bolometric luminosity, $\lbol$, of the
star, as opposed to variation in $\teff$.  However, for the very blue
stars, the bolometric corrections make it difficult to separate
changes in $\lbol$ from $\teff$, since changes in the latter result in
very small changes in color, see Figure \ref{fig:mist}.  This point
highlights the need for simultaneous monitoring of these hot stars in
the UV and optical, which would enable measurement of variability in
$\teff$ and $\lbol$.

There are 19 stars in our sample that are brighter than
$M_{\rm I,814}=-10$.  These objects are unlikely to be mis-identified
compact star clusters, but they might be unresolved binaries, in which
case our luminosities would be over-estimated by at most a factor of
two.  As discussed in Section \ref{s:cmds}, stars with
$M_{\rm I,814}<-10$ are all {\it evolved} high-mass ($>40\Msun$)
stars.  We show all 19 light curves in Figure \ref{fig:bright}.  In
each panel we plot the $\I-$band light curves within the Cycle 24 data
in the bottom panels and the 22 year baseline light curve including
the Cycle 4 and 13 data in the upper panels.  In all nine of the light
curves in the right panels there is clear variability on month, year,
and decade timescales.  In contrast, the ten variables on the left
show little evidence for variability, on both month-year and decade
timescales.  From these brightest stars we can already conclude that
variability is very common (at least 9/19 or $\approx50$\%) amongst
the luminous star population.

In Appendix \ref{s:exlc} we present additional light curves of various
stellar types, including phase-folded light curves of periodic
variables.

\begin{figure*}[!t]
\center
\includegraphics[width=0.9\textwidth]{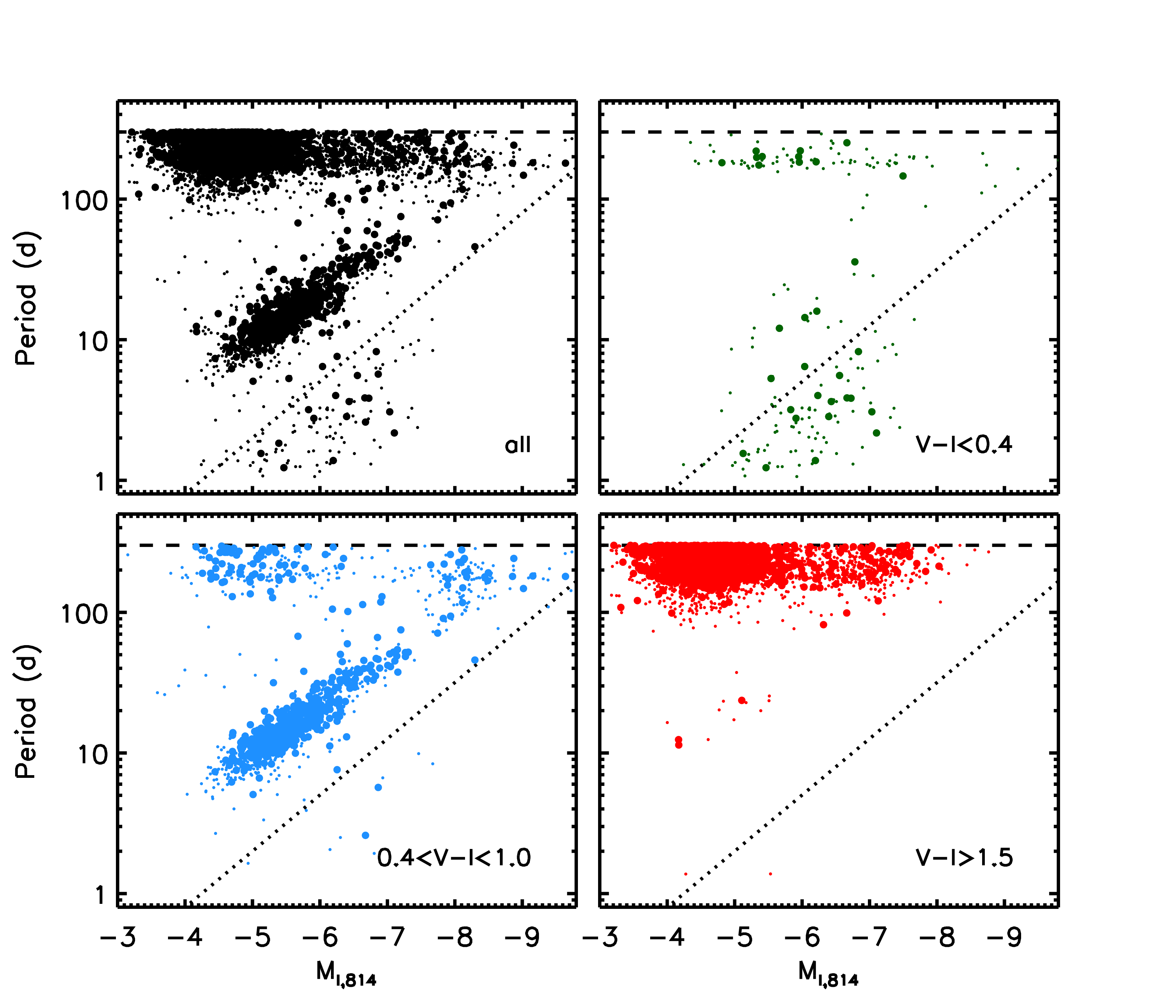}
\vspace{0.1cm}
\caption{Period as a function of magnitude for all stars with
  significant detected periods.  Stars with Lomb-Scargle power $>13$
  are shown as large symbols; small symbols include stars with
  power$>10$. The overall sample of periodic variables is shown in the
  upper left, and the remaining panels show the sample split according
  to $\V-\I$ color.  The dashed line is our nominal detection limit
  owing to the baseline of the Cycle 24 data, and the dotted line is
  meant to guide the eye.  Inspection of the light curves for stars
  below the dotted line reveals that many are eclipsing binaries.
  Periods in excess of $\approx100$ d should be interpreted with
  caution; visual inspection reveals that the light curves contain
  smooth variation on long timescales but the exact timescale is
  difficult to determine with our data.}
\label{fig:periodmag}
\end{figure*}

\subsubsection{The variable star population across luminosity and
  color}

We now turn to a quantitative analysis of the light curve variation
for our overall sample.  We begin by considering the light curve rms
measured in the $\I-$band, $\Irms$, as a measure of variability.  We
plot this quantity vs. the absolute $\I-$band magnitude,
$M_{\rm I,814}$, in Figure \ref{fig:vfrac1}.  We also show the
variability threshold ($\Inorm>2.0$) as a grey line.  The plume of
stars at $M_{\rm I,814}>-5$ and with large $\Irms$ are AGB stars with
Mira and SRV-like variability.  An important result from this work is
that for luminous stars, e.g., $M_{\rm I,814}<-6$, the amplitude of
variability is apparently continuous, rather than there being distinct
``non-variable'' and strongly variable classes.  Stars with
$M_{\rm I,814}<-7$ are shown as larger symbols and are color-coded by
their $\V-\I$ color.  Notice that the overwhelming majority of red
stars lie above the variability threshold.

In Figure \ref{fig:varvi} we compare the $\V$, $\I$, and $V-I$ light
curve rms values for a bright subsample of stars.  We chose a narrow
range of bright magnitudes for this comparison so that the sample has
a small and narrow range of photometric uncertainties, thereby
enabling a straightforward interpretation of the computed rms values.
We see clearly that $\Vrms$ and $\Irms$ are strongly correlated and
there is slightly larger variance in $\V$ compared to $\I$.  In
contrast, $(V-I)_{\rm rms}$ is generally very low and mostly
consistent with the photometric uncertainties (the median of the
quadrature sum of the $\V$ and $\I$ band uncertainties is shown as a
horizontal line in the bottom panel).  These results imply that the
light curves are varying coherently in the $\V$ and $\I$ bands.  See
Appendix \ref{s:exlc} for examples.

Figure \ref{fig:vfraccmd} shows the variable star fraction across the
CMD.  The variable fraction was computed in small color-magnitude
bins; at the bright end this fraction is noisy due to Poisson
fluctuations.  For this reason, in the bottom panels we show the
variable star fraction as a function of color in larger bins of
luminosity (left panel) and as a function of luminosity in larger bins
of color (right panel).

There are many interesting features in this figure.  One clearly sees
the Cepheid instability strip at $\V-\I\approx0.7$, as well as the
supergiant variables (e.g., Mira and SRV variables) at red colors.  It
is interesting that the variability fraction in instability strip is
never very high, although it is higher than both bluer and redder
colors at fixed magnitude.  There appears to be two populations of
variables at red colors, which is also apparent in the lower right
panel.  This is likely due to RSGs at the bright end and AGB stars at
the fainter end.  One also sees a high fraction of variability amongst
the brightest blue stars.  At least some of these are likely LBV
candidates.  We remind the reader that at $M_{\rm I,814}<-7$
essentially every bright blue star is an {\it evolved} massive star
(see Figure \ref{fig:mist}).

We also show in this figure a reddening vector whose length
corresponds to $E(B-V)=0.2$, which we regard as a typical value for
internal reddening within a metal-rich spiral galaxy.  In a face-on
disk galaxy like M51, where most of the dust and gas are confined to
the mid-plane, we can expect roughly half of the older stars (in front
of the disk) to experience very little attenuation, and roughly half
(behind the disk) to experience significant attenuation.  For young
stars still embedded in their natal clouds we might expect overall
higher levels of extinction.  It is beyond the scope of this paper to
attempt to model these effects; we simply note them here as possible
complicating factors when interpreting the observed CMDs.  See e.g.,
\citet{Dalcanton15} for a sophisticated treatment of dust extinction
in a spiral galaxy (M31).

We also show in the upper panel of Figure \ref{fig:vfraccmd} regions
where theoretical models predict variability.  The instability strip
is estimated from \citep{Paxton15} with a width meant to approximate
both our data and the LMC Cepheids from OGLE \citep{Soszynski15}.  The
supergiant instability region is estimated by applying the pulsation
growth rate equation from \citet{Yoon10} to our MIST stellar tracks
and marking the region where the growth rate is $>1$.  \citet{Yoon10}
find that only stars with $M>15\Msun$ pulsate, and they only consider
stars up to $40\Msun$.  Here we mark stars with RSG instabilities in
the mass range $15<M<40\Msun$ as solid red lines and $10<M<15\Msun$ as
dashed lines.  Notice that instabilities at lower masses seem to be
required to reproduce the observed region of variability in our data.

We also consider regions where the fundamental pulsation mode of an
evolved giant would have a period $>100$ d; this is indicated by the
orange hatched region.  For the cool stars we have included the
effects of circumstellar dust in the model predictions following
\citet{Villaume15}.  This has a relatively minor effect on the
observed CMD except for the lower-mass AGB stars, which, having
intrinsically lower $\teff$, tend to produce large quantities of dust
in their final stages of evolution.

\begin{figure*}[!t]
\center
\includegraphics[width=0.9\textwidth]{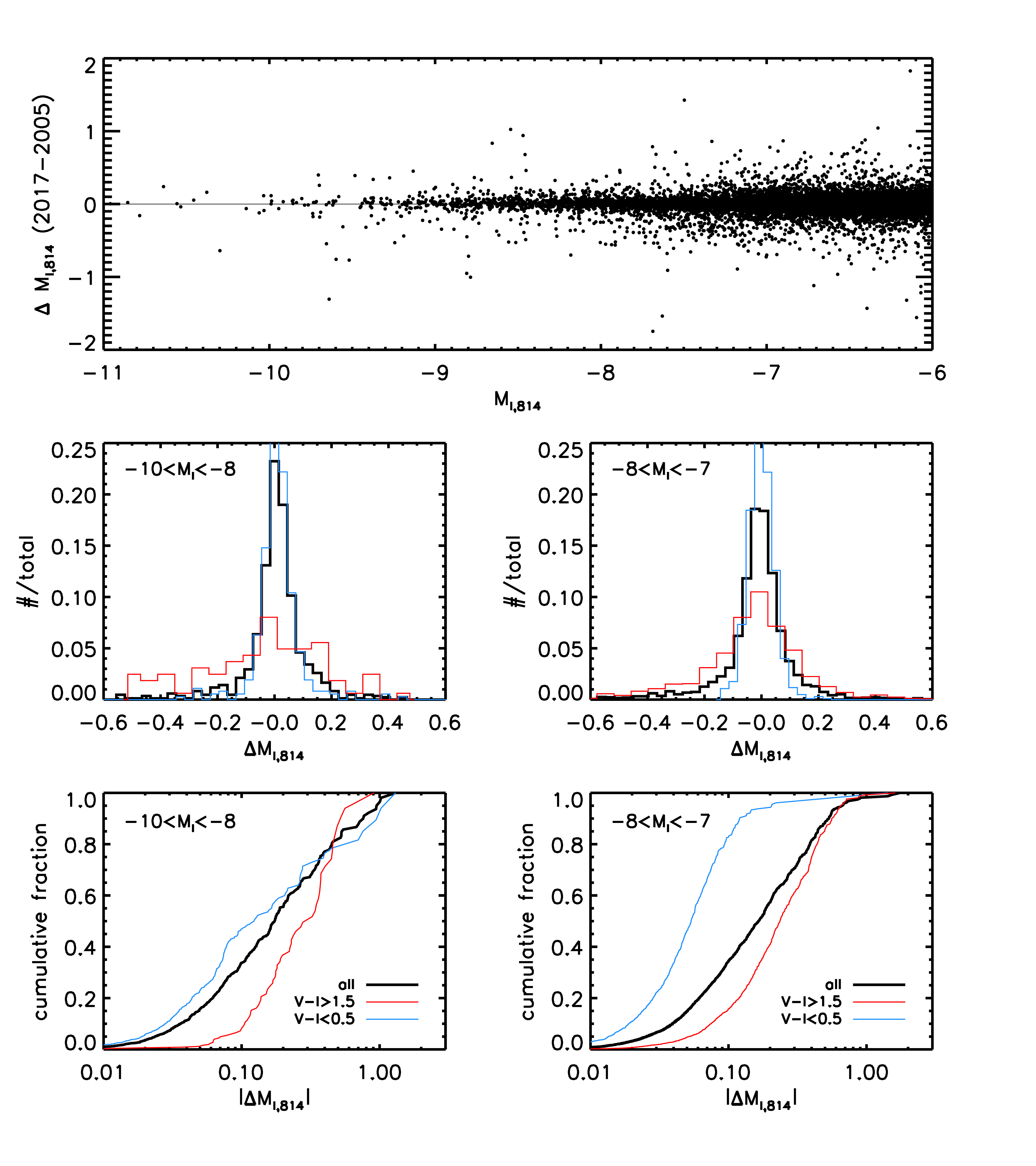}
\caption{Variability measured on a 12 year timescale between 2005 and
  2017 for stars with crowding parameter $<0.1$.  {\it Top Panel:}
  difference in $\I-$band magnitude as a function of $\I-$band
  magnitude in 2017.  {\it Middle Panels:} Differential distribution
  of $\Delta M_{\rm I,814}$ in two magnitude bins.  Results are shown
  both for all stars within the bin and stars separated by $V-I$ color
  (avoiding the instability strip at $\V-\I\approx0.7$).  The
  histogram bin width for the red selection is twice that of the other
  selections due to the fewer number of red stars. {\it Bottom
    Panels:} Cumulative distribution of $\I-$band magnitude change
  separated by color.  The lower left panel shows stars with
  $-10<M_{\rm I,814}<-8$ while lower right shows stars with
  $-8<M_{\rm I,814}<7$.}
\label{fig:deltamag10}
\end{figure*}

\begin{figure*}[!t]
\center
\includegraphics[width=0.9\textwidth]{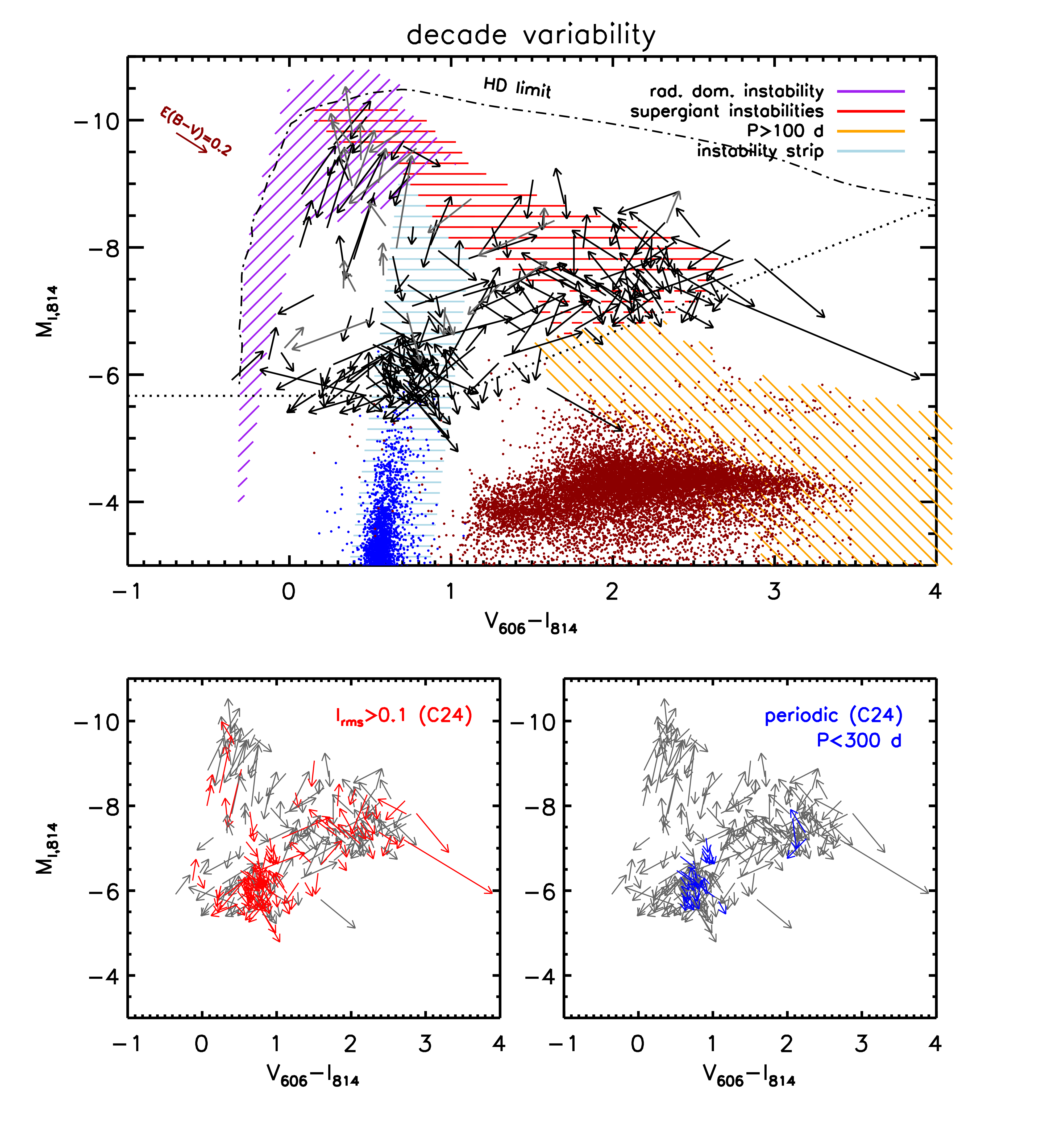}
\vspace{0.1cm}
\caption{Change in color and magnitude for the most extreme variables
  measured on decade timescales.  Variables were selected to have
  $\Delta m>0.3$ between 2005 and 2017 (black arrows) or between 1995
  and 2005 (grey arrows).  Notice that several arrows are connected
  indicating that the same star is varying over these two intervals.
  Also shown is the Humphreys-Davidson (HD) limit (dot-dashed line)
  and Cepheids, Mira and SRV variables from the LMC (blue and dark red
  symbols).  Hatched regions indicate where radiation-dominated
  related instabilities (purple region), red supergiant instabilities
  (red region), fundamental mode pulsation periods $>100$ d (orange
  region), and the instability strip (light blue region) are expected
  to occur.  Stars in these regions are likely to vary at some level,
  and indeed most of the large amplitude variables in M51 reside
  within these regions.  A reddening vector for $E(B-V)=0.2$ is also
  shown.  Lower panels show the same arrows now marked by whether they
  show $\Irms>0.1$ in the Cycle 24 data (left panel) or are periodic,
  with a Lomb-Scargle periodogram power$>13$ and period $<300$ d in
  the Cycle 24 data (right panel).}
\label{fig:arrows}
\end{figure*}

Finally, we consider instabilities that may arise from
radiation-dominated envelopes in massive stars \citep[e.g.,][and
references therein]{Paxton13, Jiang15, Owocki15}.  In evolved massive
stars conditions can arise in which a portion of the envelope is some
combination of radiation-dominated ($P_{\rm gas}/P\approx0$),
approaching the local Eddington limit, and experiencing a density
inversion.  The ultimate cause of these conditions is believed to be
the iron opacity bumps at $T\approx10^{5.3}$ K and $10^{6.3}$ K.  It
is difficult or perhaps impossible to properly model this phase of
stellar evolution in 1D, but our goal here is simply to use our 1D
MIST models to identify periods in the evolution of massive stars when
these conditions arise \citep[see][for examples of 3D simulations
relevant to this phenomenon]{Jiang15, Jiang17}.  Specifically, we
identify the region in the CMD where the MIST models satisfy
\texttt{gradt\_excess\_alpha}$>0.5$, where this internal MESA variable
is a function of two additional variables,
$\lambda_{\rm max} \equiv {\rm max} (L_{\rm rad}/L_{\rm Edd})$ and
$\beta_{\rm min}\equiv {\rm min} (P_{\rm gas}/P)$; see
\citet{Paxton13} for details.  Our particular choice of 0.5 is
somewhat arbitrary and the particular value will change the exact
boundary of the purple hatched region in Figure \ref{fig:vfraccmd}.
Our goal is merely to illustrate where stars in this phase are
expected to lie in CMD space.

Overall, the broad agreement between the expected and observed regions
of instability is encouraging.  However, the correspondance is not
perfect.  There are blue variables at fainter magnitudes than
predicted by our simple prescription for identifying radiation
dominated envelopes of massive stars.  At cool temperatures, the
predicted lower-mass limit of RSG instabilities of $15\Msun$ by
\citet{Yoon10} is unable to account for the lower luminosity variables
in our sample.  Extending their instability criterion to $10\Msun$
results in better agreement with the data.  The models are not able to
reproduce the colors of the coolest variables.  We have adopted a
simple prescription for circumstellar dust around these evolved stars;
more detailed modeling may be required for these very cool stars.  The
onset of fundamental mode pulsations with $P>100$ d does a good job of
reproducing the location of the lower-luminosity cool variables,
although the data include a population of warmer variables at
$1<\V-\I<2$ that are not predicted by (solar metallicity) fundamental
mode pulsations with $P>100$ d.  We leave a more detailed comparison
of the observed and theoretical variability regions for future work.

The bottom panels of Figure \ref{fig:vfraccmd} shows the variability
fraction, as a function of luminosity and color.  The variability
fraction is a strongly increasing function of both luminosity and
color with interesting non-monotonic behavior at the color of the
Cepheid instability strip (bottom left panel) and at
$M_{\rm I,814}\approx-6$ amongst the redder stars (bottom right panel).
The variability fraction exceeds 50\% for RSGs, which are red and
brighter than $M_{\rm I,814}\approx-7$.  This variability fraction is
in broad agreement with recent work in M31, where \citet{Soraisam18}
found that all RSGs brighter than $M_K=-10$ show evidence for
variability (the typical RSG color is $I-K\approx3$).  Amongst the
bluest stars, the variability fraction increases to $\approx20$\% for
the brightest stars.  

We emphasize that the variability fractions quoted here are lower
limits.  As noted in previous sections, our data are not sensitive to
variability at levels below $\Irms\approx0.03$.  Moreover, our
threshold for variability is $\Inorm>2.0$, which is the
error-normalized rms.  At fainter magnitudes where the errors are
larger the amplitude must therefore be larger to pass this variability
threshold.  One sees this selection effect clearly for the Cepheids in
Figure \ref{fig:vfraccmd} --- we are able to identify Cepheids as
faint as $M_{\rm I,814}=-4$ in our data (see below), but they do not
show up here due to the variability criterion.

We now consider variables that show periodic behavior as indicated by
the Lomb-Scargle periodogram analysis.  In Figure \ref{fig:periodmag}
we show all variables with $\Inorm>2.0$ and a Lomb-Scargle power $>10$
with periods of $1-300$ d, separated according to $\V-\I$ color.
Stars with power$>13$ are shown as large symbols.  As discussed in
Section \ref{s:quant}, these selections on the maximum power in the
light curve do not necessarily identify a pure nor complete sample,
but visual inspection has lead us to conclude that these values
identify large fractions of the underlying variables with low levels
of contamination.  We also emphasize that the stars with periods
$>100$ d should be interpreted with caution --- it is difficult if not
impossible to reliably measure such long periods with data spanning
only one year.  However, visual inspection reveals that such stars
{\it are} clearly variable on long timescales; see Figure
\ref{fig:exper} for examples.  We therefore recommend interpreting
stars with $>100$ d periods as showing clear signs of long timescale
variability but with periods that are poorly determined.

Several distinct classes of objects are seen in this figure.  The
Cepheid Leavitt Law shows up clearly at $0.4<\V-\I<1.0$.  The dotted
line is meant to guide the eye --- based on visual inspection, stars
below this line appear to be eclipsing binaries.  The red stars are
almost exclusively of long periods, except for a few stars at periods
of $10-30$ d which upon inspection are highly reddened Cepheids.
There is a curious population of long period ($>100$ d) very blue
stars, although the light curves in these cases are relatively erratic
and the derived periods are questionable.  Given the relatively short
baseline of our Cycle 24 data and the often somewhat irregular (or at
least not perfectly sinusoidal) light curves, it is difficult to
finely resolve the various classes of long period variables seen in
longer baseline data \citep[e.g.,][]{Soszynski09}.

\subsection{Variability on decade timescales}

The archival \HST\, data obtained in 1995 and 2005 allow us to probe
stellar variability on decade timescales amongst 18,000 stars brighter
than $M_{\rm I,814}<-6$ in both datasets.  The 2005 data are both
substantially deeper and of higher quality (ACS vs. WFPC2) and so most
of the discussion in this section focuses on comparing our Cycle 24
data to the 2005 data.

In Figure \ref{fig:deltamag10} we plot the change in $\I-$band
magnitude between the Cycle 24 data (referred to as the `2017' data
throughout this section) and the 2005 archival data.  The upper panel
shows $\Delta M_{\rm I,814}$ as a function of $M_{\rm I,814}$, the
middle panels show and the differential distribution of
$\Delta M_{\rm I,814}$ separated by color, and the lower panels show
the cumulative distribution of $|\Delta M_{\rm I,814}|$, also split
according to the Cycle 24 $\V-\I$ color.  In the lower sets of panels
the data are separated into two magnitude bins.  The distribution of
variability amplitudes in the top panel is consistent with being drawn
from a single function of the form $N\propto(|\Delta M|)^{-1.5}$.
This suggests that the large amplitude variables are simply the tail
of a distribution rather than being a special class of objects.

At $-10<M_{\rm I,814}<-8$ the distribution shows a weak dependence on
color and implies that $\approx20$\% of such stars have varied by
$>0.5$ mag over 12 years and $\approx70$\% have varied by $>0.1$ mag
over the same interval.  By $-8<M_{\rm I,814}<-7$ there is a much
stronger color-dependence such that only $\approx10$\% of blue stars
varies by more than 0.1 mag, compared to 80\% of red stars.
Variability exceeding 1 mag is extremely rare, occurring in only 14
stars out of the 18,000 brighter than $M_{\rm I,814}<-6$.  Our Cycle
24 data covers 40\% of the $\I-$band flux of M51 and so we can conclude
that a typical $L^\ast$ metal-rich star-forming galaxy contains
$\sim30-40$ stars brighter than $M_{\rm I,814}<-6$ that vary by more
than one mag on decade timescales.

In Figure \ref{fig:arrows} we show the change in color and magnitude
on decade timescales.  In this figure it was necessary to include
several additional cuts on the data, owing to the varying depths of
the Cycle 4 and 13 data.  For the Cycle 13 data we require $I<24$ and
magnitude errors of $<0.1$ in both filters.  This yields an
approximate color-dependent lower limit indicated by the dotted line
in the upper panel.  We require similar magnitude precision for the
Cycle 4 data, which results in even effective brighter limits owing to
the short exposure time of those data.  We select variable stars as
those with $\Delta m>0.3$ in either the 1995-2005 or 2005-2017
intervals.  In the upper panel, black arrows show change from 2005 to
2017 and grey arrows show change from 1995-2005.  Notice that in some
cases the grey and black arrows are connected, indicating that a
single star has varied significantly in both intervals.

In this figure we also indicate regions where stellar variability is
expected on theoretical grounds (see also Figure \ref{fig:vfraccmd}),
and the Humphreys-Davidson limit \citep{Humphreys79}, which is an
empirical luminosity boundary determined from stars in the Galaxy and
Magellanic Clouds.  We also show the locations of Mira and SRV
variables from the LMC\citep{Soszynski09} and LMC Cepheids
\citep{Soszynski15} for reference.  Notice that the sample of LMC
variables terminates around $M_{\rm I,814}\approx=-6$, likely owing to
the fact that the LMC is a much lower mass galaxy compared to M51 and
so lacks these rare, bright stars.

In the lower panels of Figure \ref{fig:arrows} we highlight those
variables that show strong and/or periodic variability within our
Cycle 24 data.  In the left panel we highlight stars with $\Irms>0.1$
and in the right panel we highlight stars that show strong evidence
for periodic behavior in their $\I-$band light curves with periods
$<300$ d.  RSGs usually have periods well in excess of 300 d
\citep[e.g.,][]{Soraisam18}, so it is not surprising that few of the
RSGs are labeled as periodic within our Cycle 24 data.  Only a
fraction of stars with large decade-scale variability also show large
amplitude variability on one year timescales.  Nonetheless, nearly all
of the stars shown as arrows in the lower left panel do harbor
statistically-significant evidence for variability within our Cycle 24
data, as defined by our metric $\Inorm>2.0$.  This result, combined
with the results of Figure \ref{fig:vfraccmd}, suggests a picture in
which nearly all luminous, evolved stars vary on year timescales,
albiet with small amplitudes, while on decade timescales these same
stars occasionally undergo much larger changes in monochromatic
fluxes.

As in Figure \ref{fig:vfraccmd}, it is striking that in Figure
\ref{fig:arrows} the majority of the decade timescale variables lie in
regions of the CMD where variability is expected on theoretical
grounds.

Finally, we have also searched for stars that have disappeared between
2005 and 2017.  There are 20,000 stars brighter than
$M_{\rm I,814}=-6$ in the 2005 data, which includes all of the RSGs
with initial masses $>10\Msun$.  All of these stars are still present
in 2017.  \citet{Kochanek08} estimate that it would require monitoring
$\sim10^6$ supergiants over a year in order to witness an event of
some kind, whether a supernova or a disappearing star (failed
supernova).  Our baseline of 12 years reduces the number of necessary
stars to $\sim10^5$, so in monitoring ``only'' 20,000 stars it is not
surprising that we have not seen a disappearing star.  If we expand
the luminosity to include the more luminous AGB stars,
$M_{\rm I,814}<-5$, we again find no disappearing stars among the
66,000 stars above this magnitude limit (and with a crowding parameter
in the 2005 data $<0.5$).


\section{Summary}
\label{s:sum}

In this paper we have presented a complete census of stellar
variability amongst the luminous star population in the face-on
$L^\ast$ spiral galaxy M51.  Our Cycle 24 data in combination with
archival Cycle 4 and 13 data enabled sensitivity to stellar
variability on day to decade timescales.  We now summarize our main
results.

\begin{itemize}

\item Stellar variability is ubiquitous.  The variability fraction is
  $\approx50$\% for stars with $M_{\rm I,814}<-7$; this fraction
  reaches $\approx100$\% for the brightest red stars (RSGs).  A wide
  variety of light curve behavior is seen amongst these bright stars,
  with smooth, long-timescale variability more common for redder
  stars, and erratic variability common for the blue stars, with
  amplitudes often exceeding one mag.  The color variation is modest
  for these large amplitude variables, suggesting at least for the
  redder stars that we are witnessing variation in $\lbol$ over time.

\item Variable stars occupy well-defined regions in the CMD.  These
  regions correspond to the instability strip, the location of RSGs
  and AGB stars (Miras and SRVs), and the locations where massive stars
  are expected to exhibit radiation-dominated and hence unstable outer
  envelopes.  There is broad qualitative agreement between observed
  and predicted locations of variables in the CMD, but there are
  several locations that require further study, including the faint
  RSGs and the luminous blue variables.

\item Stars experiencing variability on {\it both} short (month-year)
  and long (decade) timescales is common but not universal.  We
  identify many cases where a star changes in brightness by $>0.3$ mag
  on a decade timescale but is approximately constant ($\Irms<0.1$) on
  month-year timescales.  This is suggestive of periods of instability
  followed by periods of quiescence, or perhaps very long timescale
  variability.

\item The amplitude of variability for luminous stars
  ($M_{\rm I,814}<-7$) on both month-year and decade timescales is
  consistent with a single, continuous distribution from small to
  large amplitudes.  In other words, there is no evidence in our data
  for distinct classes of large and small amplitude variables amongst
  the most luminous stars.  This suggests for example that LBVs are
  the extreme end of a continuum of variability amplitudes.

\end{itemize}

A key missing piece to our analysis of variability in M51 is physical
parameters for the stars, especially $\teff$ and $\lbol$.  Archival
$B$ through $H-$band data do exist and overlap with the footprint of
our Cycle 24 data, but they were taken at different epochs, which
makes them less useful for estimating stellar parameters of variable
stars.  Moreover, for the hottest stars, UV data is essential for
accurate stellar parameters.  An important next step is to acquire
simultaneous UV--NIR photometry in order to measure accurate $\teff$,
$\lbol$, and reddening values on a star-by-star basis.

In the near future we can expect similar censuses to become available
in M31 (via PTF data) and in the Galaxy (via Gaia data).  Large
all-sky monitoring efforts currently planned and underway, including
the Zwicky Transient Factory (ZTF) and the Large Survey Synoptic
Telescope (LSST), will enable the measurement of stellar variability
throughout the Local Group.  These studies of variability will offer
powerful and novel probes of stellar interiors and short lived phases
of stellar evolution, and will require an equally concerted effort
from stellar modelers in order to bright to light the basic (and not
so basic) physical processes governing stellar variability.


\acknowledgments 

We thank Nathan Smith for very helpful comments on an earlier version
of the manuscript.  CC and JS acknowledge support from the Packard
foundation; CC, BJ, JS, and PvD acknowledge support from NASA grants
HST-GO-14704.001, HST-GO-14704.002, and HST-GO-14704.003.  DRW is
supported by a fellowship from the Alfred P. Sloan Foundation.  CC
acknowledges the Sorg Chateau where this paper was written.


\begin{appendix}

\section{Example light curves}
\label{s:exlc}

In this Appendix we provide additional example light curves associated
with several figures in the main text.

\begin{figure*}[!t]
\center
\includegraphics[width=0.95\textwidth]{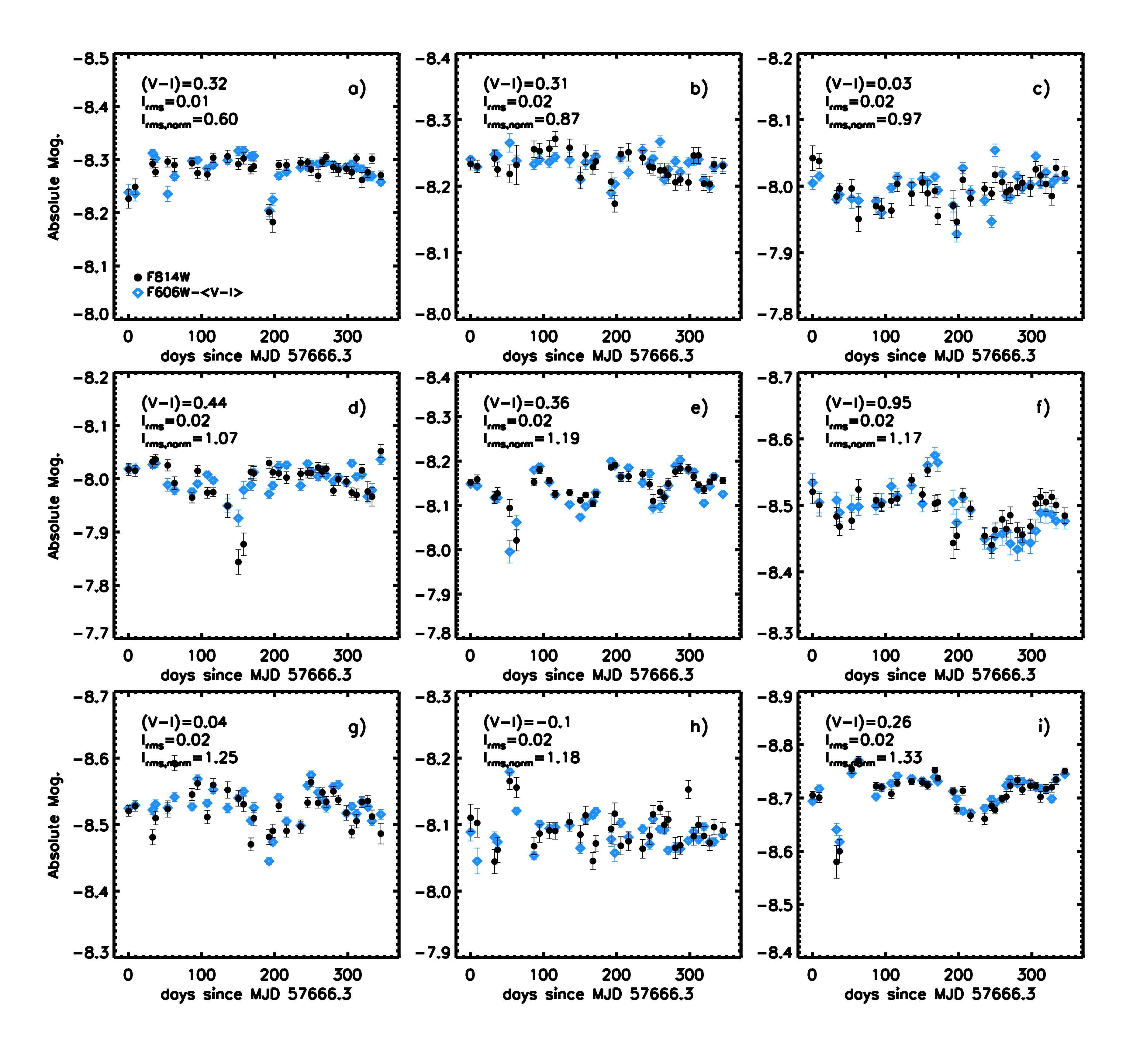}
\vspace{0.1cm}
\caption{Randomly drawn light curves for stars from Figure
  \ref{fig:vfrac1} with $-9<M_{\rm I,814}<-8$, sorted by $\Irms$.}
\label{fig:brightlc}
\end{figure*}

Figure \ref{fig:vfrac1} showed the $\Irms$ vs. $M_{\rm I,814}$.  In
Figure \ref{fig:brightlc} we show a set of light curves of bright
stars from Figure \ref{fig:vfrac1}.  These are randomly drawn from
$-9<M_{\rm I,814}<-8$ and sorted by $\Irms$.  It is clear from these
examples that the variability is real $\Irms\ge0.03$.  From inspection
of all the light curves for the brightest stars, including those not
shown in Figure \ref{fig:brightlc}, it is also clear that at least
some of the stars with $\Irms<0.03$ are also true variables, so we
believe that the quoted variability fraction at the bright end is
indeed a lower limit.

\begin{figure*}[!t]
\center
\includegraphics[width=0.95\textwidth]{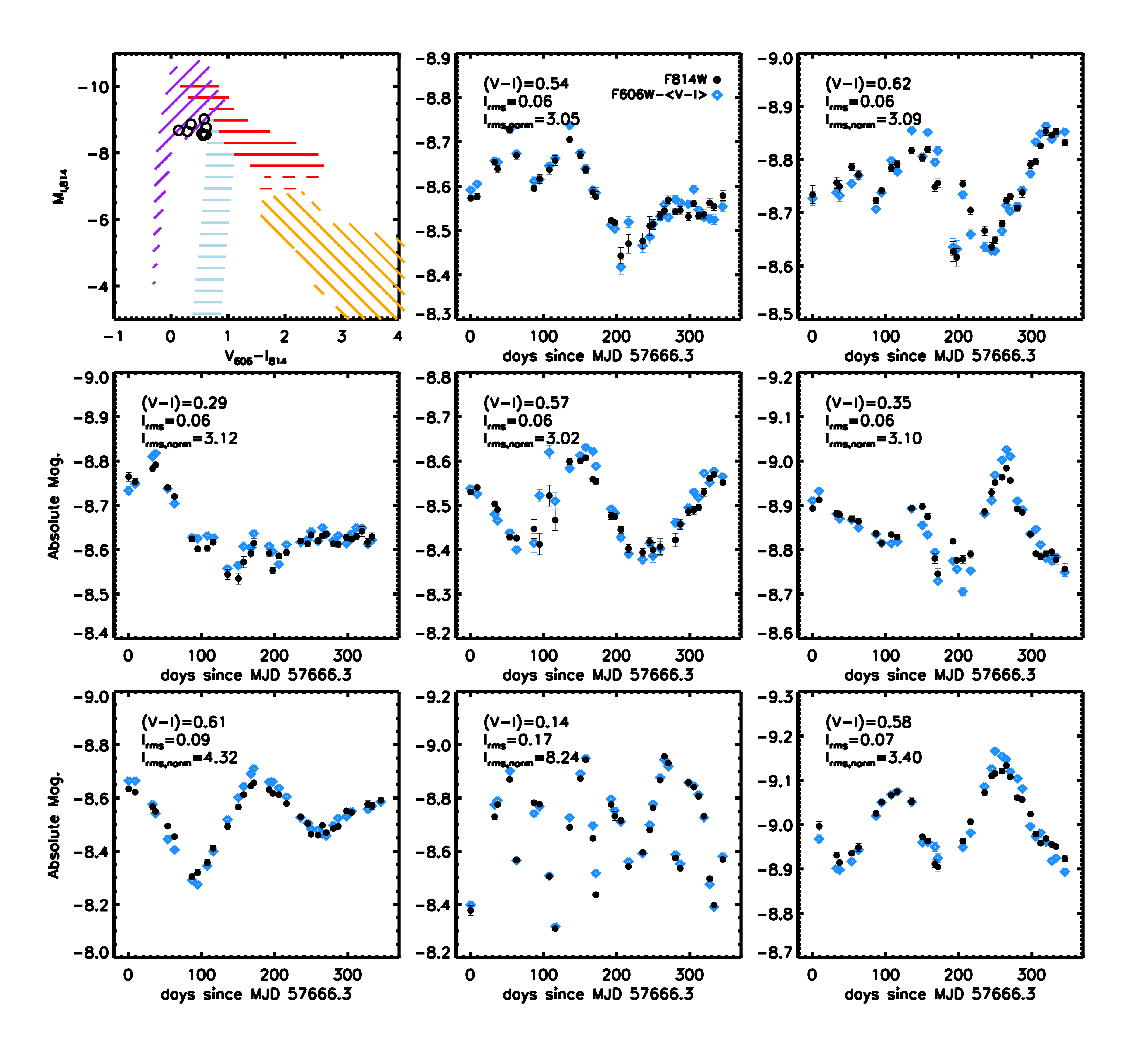}
\vspace{0.1cm}
\caption{Randomly drawn light curves for stars from the purple hatched
  region of Figure \ref{fig:vfraccmd}.  The locations of the stars in
  the CMD are shown in the upper left panel.}
\label{fig:exlbv}
\end{figure*}

\begin{figure*}[!t]
\center
\includegraphics[width=0.95\textwidth]{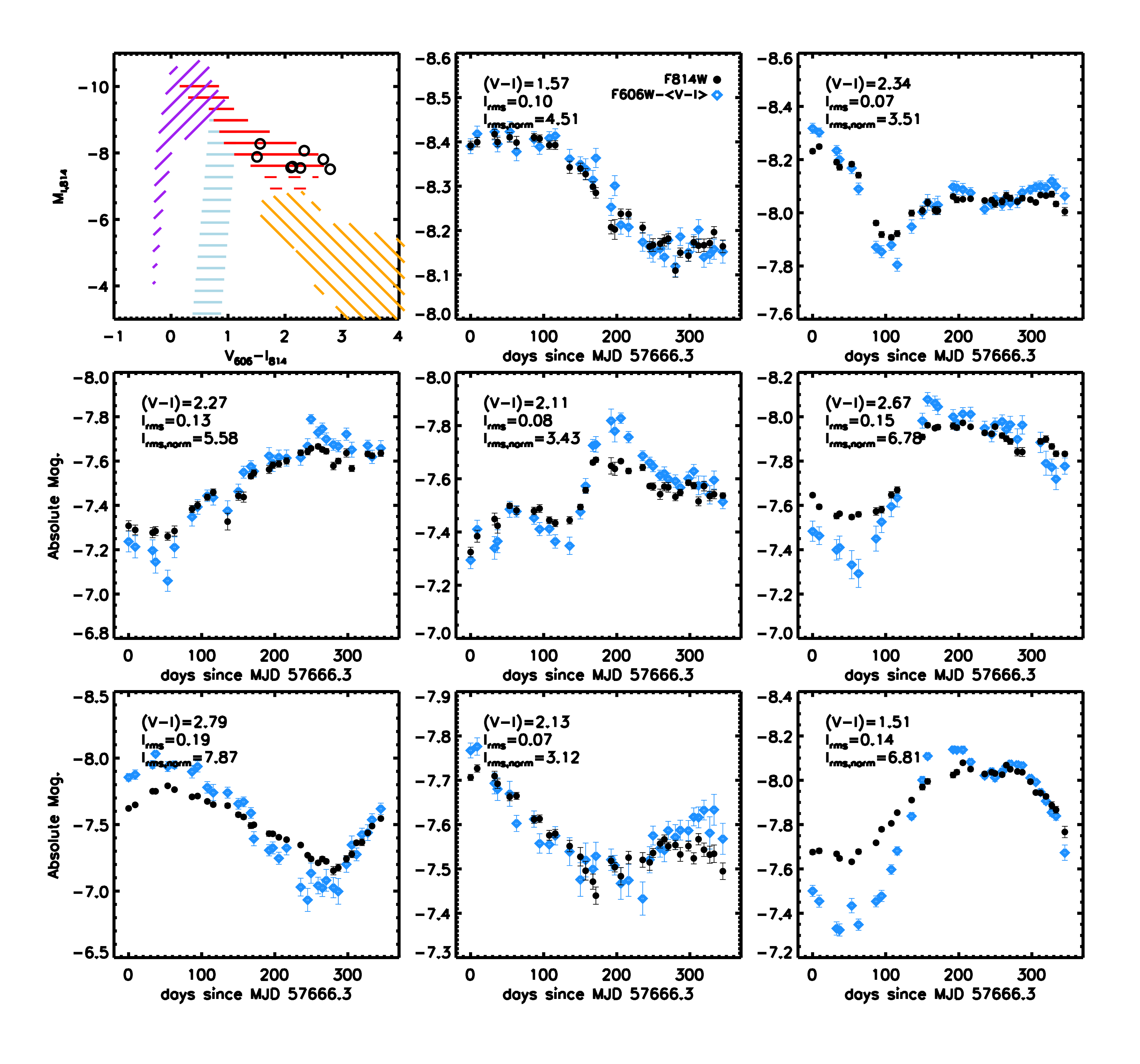}
\vspace{0.1cm}
\caption{Randomly drawn light curves for stars from the red hatched
  region of Figure \ref{fig:vfraccmd}. The locations of the stars in
  the CMD are shown in the upper left panel.}
\label{fig:exrsg}
\end{figure*}

\begin{figure*}[!t]
\center
\includegraphics[width=0.95\textwidth]{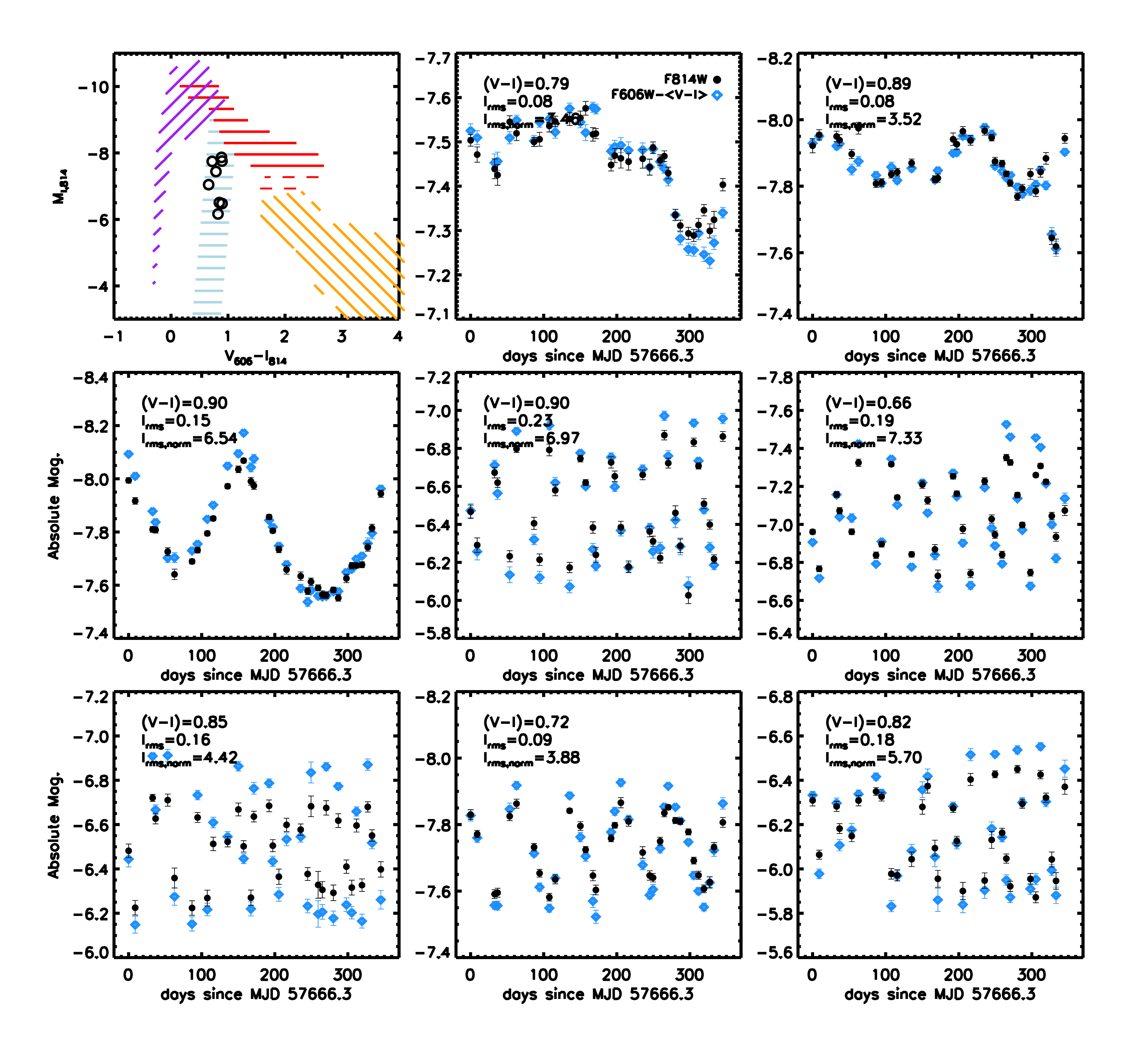}
\vspace{0.1cm}
\caption{Randomly drawn light curves for stars from the light blue hatched
  region of Figure \ref{fig:vfraccmd}. The locations of the stars in
  the CMD are shown in the upper left panel.}
\label{fig:exceph}
\end{figure*}

\begin{figure*}[!t]
\center
\includegraphics[width=0.95\textwidth]{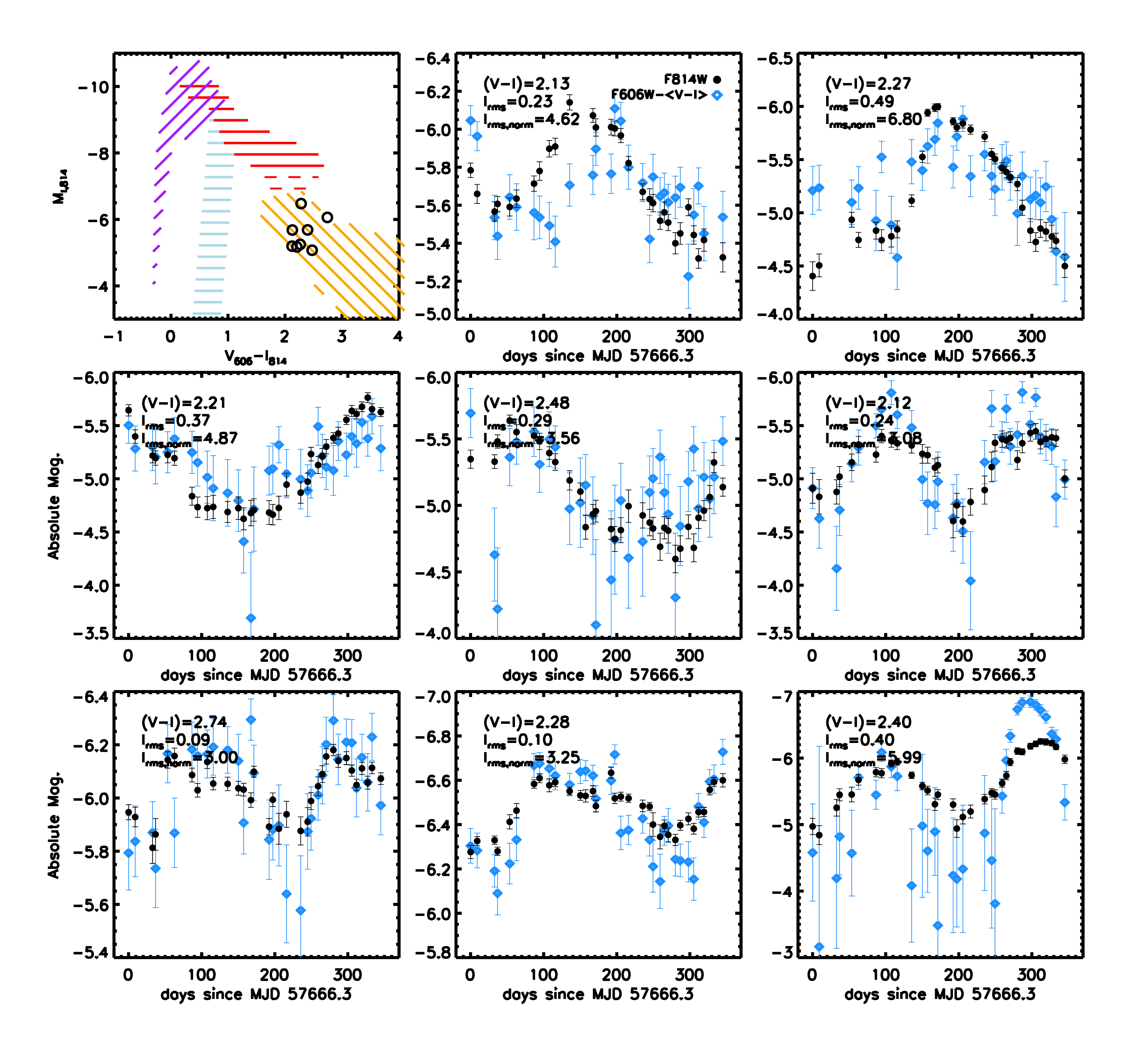}
\vspace{0.1cm}
\caption{Randomly drawn light curves for stars from the orange hatched
  region of Figure \ref{fig:vfraccmd}.  The locations of the stars in
  the CMD are shown in the upper left panel.}
\label{fig:exmira}
\end{figure*}

In Figures \ref{fig:exlbv}--\ref{fig:exmira} we present example
light curves of the variable stars from various regions of the CMD
shown in Figure \ref{fig:vfraccmd}.  In each figure we highlight stars
in one of the four hatched regions of instability. 

\begin{figure*}[!t]
\center
\includegraphics[width=0.95\textwidth]{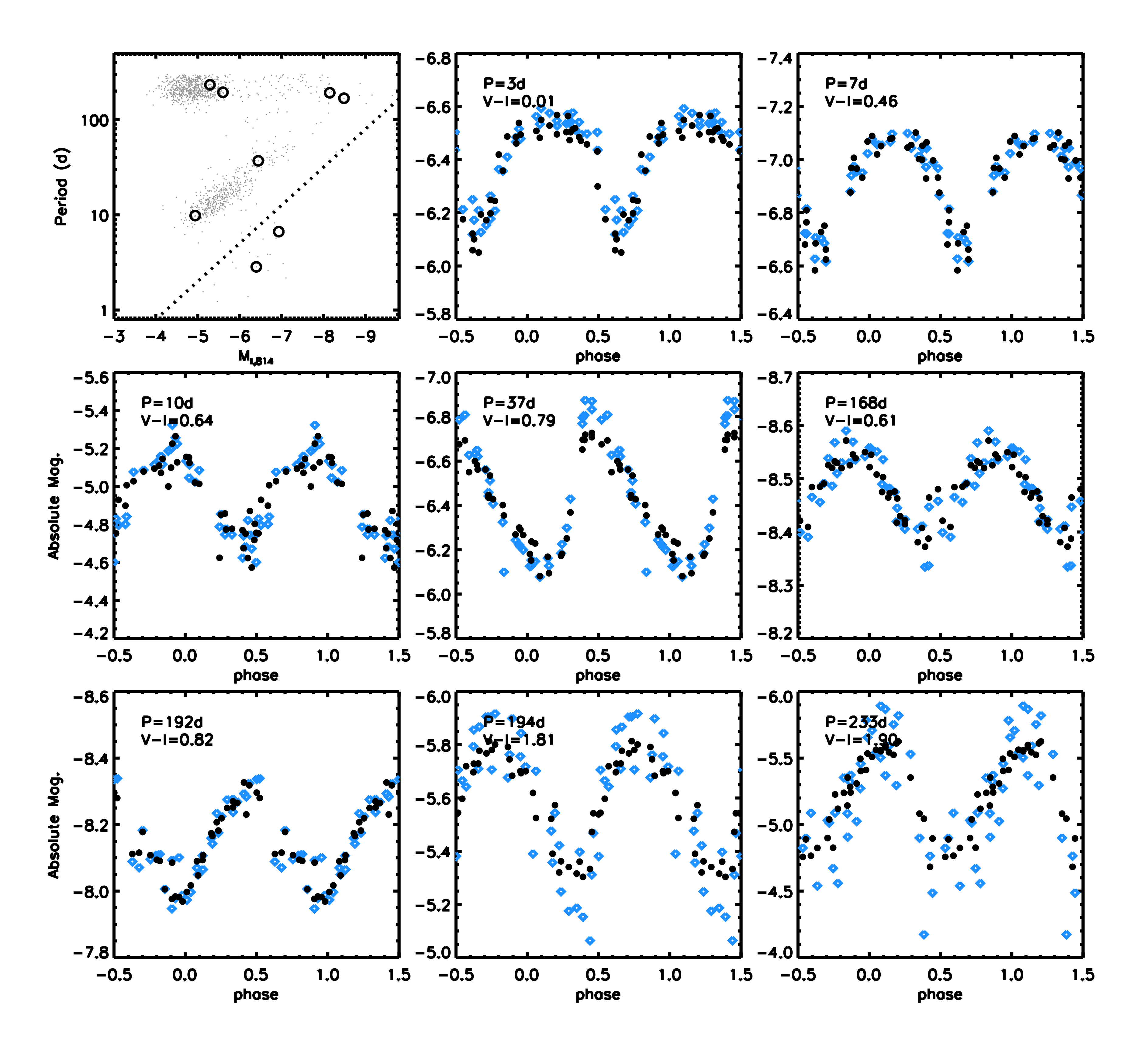}
\vspace{0.1cm}
\caption{Randomly drawn phase-folded light curves for stars selected
  from several regions of period-luminosity space.  The locations of
  the stars in Figure \ref{fig:periodmag} are shown in the upper left
  panel.  Stars are sorted by period.  The shortest periods are likely
  eclipsing binaries.}
\label{fig:exper}
\end{figure*}

In Figure \ref{fig:exper} we show phase-folded light curves of
periodic variables drawn from several locations in the
period-luminosity plane (see Figure \ref{fig:periodmag}).  The
shortest period variables are likely eclipsing binaries.

\section{Photometry and Derived Data Products}

In this Appendix we provide an overview of the fits binary file
containing the catalog of stars and derived parameters used in this
paper.  As noted in the main text, the catalog includes all stars with
a DOLPHOT object type$=1$ (star), $\I-$band crowding $<0.5$, $\I-$band
sharpness, $-0.2<$sharp$<0.2$, $\chi_I<2$, a final $\I-$band S/N$>5$
and detections in the $\I-$band for least five epochs.  We include all
stars with $M_{\rm I,814}<-5.0$, resulting in 72,623 entries.  Table
\ref{t:data} provides a description of the fits file contents.

\begin{deluxetable*}{rll}
\tablecaption{Description of Datafile}
\tablehead{ \colhead{column} & \colhead{label} & \colhead{description} }
\startdata
  1 &            RA &  Right Ascension in decimal degrees (J2000) \\
  2 &           DEC &  Declination in decimal degrees (J2000) \\
  3 &     MAGV0\_C24 &  $\V$ magnitude of combined Cycle 24 data \\
  4 &     ERRV0\_C24 &  $\V$ error of combined Cycle 24 data\\
  5 &     CHIV0\_C24 &  $\V$ $\chi$ parameter of combined Cycle 24 data\\
  6 &   SHARPV0\_C24 &  $\V$ sharpness parameter of combined Cycle 24 data\\
  7 &   CROWDV0\_C24 &  $\V$ crowding parameter of combined Cycle 24 data\\
  8 &     MAGI0\_C24 &  $\I$ magnitude of combined Cycle 24 data\\
  9 &     ERRI0\_C24 &  $\I$ error of combined Cycle 24 data\\
 10 &     CHII0\_C24 &  $\I$ $\chi$ parameter of combined Cycle 24 data\\
 11 &   SHARPI0\_C24 &  $\I$ sharpness parameter of combined Cycle 24 data\\
 12 &   CROWDI0\_C24 &  $\I$ crowding parameter of combined Cycle 24 data\\
 13 &      MAGV\_C24 &  Individual epoch $\V$ magnitudes for Cycle
 24 data\\
 14 &      ERRV\_C24 & Individual epoch $\V$ errors for Cycle
 24 data \\
 15 &      MAGI\_C24 &  Individual epoch $\I$ magnitudes for Cycle
 24 data\\
 16 &      ERRI\_C24 &  Individual epoch $\I$ errors for Cycle
 24 data\\
 17 &        MAG\_HH &  $V_{\rm 505}$ and $\I$ magnitudes of 2005 data\\
 18 &        ERR\_HH &  $V_{\rm 505}$ and $\I$ errors of 2005 data\\
 19 &        CHI\_HH &  $V_{\rm 505}$ and $\I$ $\chi$ parameters of 2005 data\\
 20 &      SHARP\_HH &  $V_{\rm 505}$ and $\I$ sharpness parameters of 2005 data\\
 21 &      CROWD\_HH &  $V_{\rm 505}$ and $\I$ crowding parameters of 2005 data\\
 22 &        MAG\_95 &  $V_{\rm 505}$ and $\I$ magnitudes of 1995 data \\
 23 &        ERR\_95 &  $V_{\rm 505}$ and $\I$ errors of 1995 data\\
 24 &        CHI\_95 &  $V_{\rm 505}$ and $\I$ $\chi$ parameters of 1995 data\\
 25 &      SHARP\_95 &  $V_{\rm 505}$ and $\I$ sharpness parameters of 1995 data\\
 26 &      CROWD\_95 &  $V_{\rm 505}$ and $\I$ crowding parameters of 1995 data\\
 27 & MATCH\_TO\_HH\_95 & Matched separation between the C24 and Archival
 catalogs (pixels) \\
 28 &  MAG\_95\_F606W &  Converted $\V$ magnitude for 1995 data \\
 29 &  MAG\_HH\_F606W &  Converted $\V$ magnitude for 2005 data \\
 30 &   LS\_PERIOD\_V &   Lomb-Scargle period at maximum power in $\V$\\
 31 & LS\_MAXPOWER\_V &  Lomb-Scargle maximum power in $\I$ \\
 32 &   LS\_PERIOD\_I &  Lomb-Scargle period at maximum power in $\I$\\
 33 & LS\_MAXPOWER\_I &  Lomb-Scargle maximum power in $\I$ \\
 34 &         RMS\_V &  rms of $\V$ light curve ($\Vrms$) \\
 35 &         RMS\_I &   rms of $\I$ light curve ($\Irms$) \\
 36 &     RMSNORM\_V &  error-normalized rms of $\V$ light curve
 ($V_{\rm rms,norm}$) \\
 37 &     RMSNORM\_I & error-normalized rms of $\I$ light curve ($\Inorm$)
\enddata
\vspace{0.1cm} 
\label{t:data}
\end{deluxetable*}

\end{appendix}



\end{document}